\documentstyle[psfig,12pt,amstex,righttag]{article}
\numberwithin{equation}{section}

\begin{document}
\begin{flushright}
hep-th/9701147 \\
January, 1997
\end{flushright}
\vspace{2cm}

\begin{center}
{\LARGE { F Theory Vacua in Four Dimensions \\  and Toric Threefolds}}
\vspace{1cm}

{\sc Kenji   \ Mohri}
\end{center}
%
%
\begin{center}
{{\it National Laboratory for High Energy Physics, Ibaraki  305 Japan} \\ 
mohri@@theory.kek.jp}
\end{center}
\vspace{5cm}

\begin{abstract}
We investigate D=4, N=1  F theory models realized by 
type IIB string compactification on toric threefolds.
Massless spectra, gauge symmetries, phase transitions 
associated with divisor contractions and flops, 
and non-perturbative superpotentials are analyzed 
using  elementary toric methods. 
\end{abstract}

\newpage
\section{Introduction}
F theory \cite{Va,MV1,MV2} provides  a remarkable way of D=4, N=1
 string compactification \cite{BS,BLS,KS,SVW,Mayr,Wi1,Wi2,KV-BJPSV}
 on a elliptic Calabi-Yau fourfold  \\
$\pi: X_{4}$ $\rightarrow $ $B_{3}$. %
In F theory, complex moduli and $\mbox{E}_{8}\times \mbox{E}_{8}$ 
gauge bundle moduli of the dual heterotic string are unified
into the parameters of $X_{4}$ \cite{MV2,sixmen}.
F theory also enables us a practical way of 
evaluating non-perturbative superpotential  \cite{Wi2,DGW,BLS,KV-BJPSV}. \\
The  aim of  this article is to provide  several examples of
D=4 F theory models realized by type IIB compactification
on toric threefolds \cite{MP,AGM,Oda,Fulton}
and to describe explicitly their massless spectra, gauge symmetries,
phase transitions and  non-perturbative superpotentials.\\
The organization of this paper is as follows:
In section 2 we describe the elements of D=4 F theory compactification.
In section 3 we treat examples of models which don't have non-Abelian gauge 
symmetry.
In particular we describe the quantum numbers 
and the blowing-up/down transition scheme of the toric Fano threefold models.
In section 4 ${\Bbb P}^{1}$ bundles over ${\Bbb P}^{2}$
models   which we call ${\Bbb G}_{n}$
are  analyzed in some detail.
Here we compute the physical Hodge numbers as well as 
the non-Abelian gauge symmetry which arises for  generic choices of the 
parameters. Then we show that the transition between 
${\Bbb G}_{n}$ models can be described as a  blowing-up/down of
 toric threefolds. 
Dual heterotic string description of this transition
including non-perturbative vacua is also presented.
In section 5 we give a example of the flop transition
which interpolate between two heterotic strings compactified on different
elliptic Calabi-Yau threefolds.
In section 6  we treat ${\Bbb P}^{2}$ bundle over ${\Bbb P}^{1}$ models.
In section 7 we deal with ${\Bbb P}^{1}$  bundle
over ${\Bbb F}_{a}$ models. Particular emphasis is placed 
on those models that  can be described as Landau-Ginzburg orbifolds.
In section 8 we consider the effect of type IIB three-brane 
wrapped around an exceptional divisor of a toric threefold.
We show that it induces a non-perturbative superpotential.
We also compute the change  in physical Hodge numbers
for a contraction of an exceptional divisor.
In appendix we show the toric data of some Fano threefolds.

\vspace{2.5cm}
While completing this work we got a preprint \cite{KLRY} via hep-th archive 
which have some overlap with our results.

\newpage
\section{ D=4 Compactification of  F Theory }
\subsection{ Toric Threefold and Elliptic Calabi-Yau Fourfold }
First we describe the construction of a toric threefold $B_{3}$ 
\cite{MP,Wi4} and the
elliptic Calabi-Yau fourfold $X_{4}$ over it \cite{BLS,Mayr,KS}
which is described as a  Weierstrass model \cite{MV1,MV2,Nak,Grassi}.\\
Let $(x_{1},...,x_{n+3})$ be the homogeneous coordinates of $B_{3}$,
$Q_{i}^{a}$ the $a$-th $U(1)$ charge of $x_{i}$,
where $a$ runs from 1 to $n$,
and 
$$
D^{a}:=\sum_{i=1}^{n+3}Q_{i}^{a}|x_{i}|^{2}-r^{a}
$$
be the $a$-th Hamiltonian.
Then the toric threefold $B_{3}$
associated with the above data is defined by the Hamiltonian quotient 
\begin{equation}
B_{3}:=\{ (x_{i})\in {\Bbb C}^{n+3}| D^{a}=0 \}/U(1)^{n}.
\end{equation}
If we define the excluded set $E(r^a)$ to be the set of the points
the ${{\Bbb C}^{*}}^{n}$-orbit of which
doesn't intersect with $\{  D^{a}=0, \quad a=1,..,n \}$, then we have another 
realization of $B_{3}$ as the holomorphic quotient:
\begin{equation}
B_{3}\cong \{ {\Bbb C}^{n+3}-E \}/{{\Bbb C}^{*}}^{n}.
\end{equation}
The total space of radii vectors $(r^a)$ such that
$\{  D^{a}=0, \quad a=1,..,n \}$ is not empty is  divided into
the K\"ahler cones according to the topology of $B_{3}$,
which constitutes  the phase diagram of the model \cite{Wi1}.\\  
We realize the elliptic fibration
$\pi: X_{4} \rightarrow B_{3}$ as  follows:
\begin{equation}
(z_{3})^{2}=(z_{2})^{3}+(z_{1})^{4}z_{2}F(b)+(z_{1})^{6}G(b),
\label{Weierstrass} %
\end{equation}
where $(z_{1},z_{2},z_{3})$ is the homogeneous coordinate of 
 ${\Bbb P}_{(1,2,3)}$ and 
the coefficients $F$ and $G$ take value in
$\Gamma(-4K_{B})$ and  $\Gamma(-6K_{B})$ respectively.
In terms of toric data, this simply means that
$a$-th $U(1)$ charges of $F(x)$ and $G(x)$ are
$ 4\sum_{i=1}^{n}Q_{i}^{a}$ and $ 6\sum_{i=1}^{n}Q_{i}^{a}$
respectively.
Note that $X_{4}$ is realized by a hypersurface in the toric fivefold
with the homogeneous coordinates
$(x_{1},..,x_{n+3};z_{1},..,z_{3})$.\\
In the type IIB picture of F theory:
\begin{equation}
\mbox{F theory on } X_{4}=
\mbox{type IIB theory on } B_{3},
\end{equation}
the axion/dilaton of type IIB theory
$\tau(b):=\phi_{\mbox{\sc r}}+i\exp(-\phi_{\mbox{\sc ns}})$  depends on
the position $b \in B_{3}$ and 
is identified with the modulus of elliptic fiber over $b$.
Then the degeneration of elliptic fibers at the discriminant locus \\
$S_{2}:=\{  b \in B_{3}| 4F(b)^3 + 27G(b)^2 = 0 \}$
is interpreted as the insertion 
 of  the seven-brane \cite{GSVY,GGP,Va} 
the  world volume of which 
is $S_{2}\times {\Bbb R}^{4}$.\\
There is a $D=8$ duality between F theory and heterotic string:
\begin{equation}
\mbox{type IIB string on } {\Bbb P}^{1} \cong
\mbox{heterotic string on } E(\tau), \label{F/hetero}
\end{equation}
where  the axion/dilaton of type IIB side at generic points
of ${\Bbb P}^{1}$ is 
identified with  the modulus $\tau$
of the elliptic curve $E$ above \cite{MV2}.
If we consider type IIB compactification on a threefold with a
${\Bbb P}^{1}$ fibration
$\pi: B_{3}\rightarrow S_{2}$,
then  the  fiberwise application of (\ref{F/hetero})
leads to  heterotic string compactified on the  elliptic fibered 
Calabi-Yau threefold $Y_{3}$ over $S_{2}$.
If a ${\Bbb P}^{1}$ 
fibration $B_{3}$ is not a genuine ${\Bbb P}^{1}$ fiber bundle,
then the dual heterotic string would be a non-perturbative one
with five-branes.
\subsection{Spectrum}
A part of the massless spectrum of F theory model in four dimensions 
may be derived from
M theory on $X_{4}$ by further compactifying %
on ${\Bbb S}^{1}$ owing to the duality:
\begin{equation}
\mbox{F theory on } X_{4}\times {\Bbb S}^{1}(R)
\cong 
\mbox{M theory on } X_{4},
\end{equation}  
where the area of the fiber torus in the right hand side 
is $1/R$. 
We can then  go to the 
Coulomb phase by giving the VEV to scalars 
of  vector multiplets in $D=3$ \cite{Va,MV1}.  
Unfortunately in M theory  model above, 
a vector multiplet is indistinguishable from
a chiral multiplet  because in three dimensions
 a vector is equivalent to a scalar.\\
Thus there would be an ambiguity in identification of the
numbers of chiral and vector multiplets.
Comparing  M theory compactification  on $X_{4}$
and type IIB theory compactification on toric $B_{3}$,
we propose the following identification of the spectrum%
\footnote{%
For any toric threefold,
cohomology groups of odd degree vanish.
If we take  a threefold which is not toric, e.g.
${\Bbb P}^{4}[d]$, $d=2,3,4$, for which $X_{4}$ becomes complete intersection
in a toric variety \cite{BLS}, we have generically
non-zero $H^{1,2}(B_3)$, which implies that we have rank $h^{1,2}(B_{3})$
vector multiplets and  $ h^{1,2}(X_{4})-h^{1,2}(B_{3})$ neutral
chiral multiplets.}
\begin{eqnarray}
\mbox{rank}(\mbox{vector multiplets})
 &=& h^{1,1}(X_{4})-(h^{1,1}(B_{3})+1) \label{vector} \\
\#(\mbox{neutral chiral multiplets}) 
&=& h^{1,1}(B_3)+h^{1,3}(X_{4})
+h^{1,2}(X_{4})
\end{eqnarray}
$H^{2,2}$ is also important because they enter into
the theory as the vacuum expectation value 
of the F/M/IIA theory 4-form field strength on $X_{4}$ \cite{BB,Mayr}.\\
The topological invariants of a Calabi-Yau fourfold are 
\begin{eqnarray}
\chi(X_{4}) &=& 4+2(h^{1,1}+h^{1,3})+h^{2,2}-4h^{1,2}\\
\chi(T^{*}) &=& -(h^{1,1}+h^{1,3})+h^{1,2}
= 8-\frac{1}{6}\chi(X_{4})  \label{complexdeform}\\
\chi(\wedge^{2} T^{*}) &=& h^{2,2}-2h^{1,2}
=12+\frac{2}{3}\chi(X_{4}).
\end{eqnarray}
From the last two  equations, we obtain a constraint on the spectrum:
\begin{equation}
 -4(h^{1,1}+h^{1,3})+2h^{1,2}+h^{2,2} = 44  \label{44},
\end{equation}
which can be used as a consistency check of
 the spectra of various models treated below.
\subsection{Spectrum of Toric Hypersurface}
When $X_{4}$ is a hypersurface in a toric fivefold
as our case and the Newton polytope $\Delta$ of which is reflexive,
the  Hodge numbers of a MPCP (maximal projective crepant partial)
 resolution of $X_{4}$,which is independent of the choice of a MPCP resolution,
are determined from the combinatoric data of the ambient toric variety
\cite{Bat,Mckay};
\begin{eqnarray}
h^{3,1} &=& l(\Delta)-6-\sum_{\dim \Theta =4}l^{0}(\Theta)
+\sum_{\dim \Theta =3}l^{0}(\Theta)l^{0}(\Theta^{*}), \nonumber \\
h^{2,1} &=& \sum_{\dim \Theta =2}l^{0}(\Theta)l^{0}(\Theta^{*}), 
\label{batyrev}\\
h^{1,1} &=& l(\Delta^{*})-6-\sum_{\dim \Theta =0}l^{0}(\Theta^{*})
+\sum_{\dim \Theta =1}l^{0}(\Theta)l^{0}(\Theta^{*}), \nonumber
\end{eqnarray}
where $\Theta$ is a face of the Newton polytope $\Delta$,
$l^{0}(\Theta)$  the number of integral points 
inside  $\Theta$, and $\Theta^{*}$  the dual face of $\Theta$
such that 
$\dim\Theta+\dim\Theta^{*}=4$.\\
%
%
It must be noted here that the reflexivity of the Newton polytope
does not assure the existence of a smooth resolution of the  Calabi-Yau
fourfold $X_{4}$.                        
A MPCP resolution $\hat{X_{4}}$ of $X_{4}$ 
may leave  terminal point singularities that  cannot be resolved
without destroying the Ricci flat condition. \\
For example 
the (non-elliptic) Calabi-Yau fourfold 
$X_{4}={\Bbb P}_{(1,1,1,1,2,2)}[8]$ \\
has four ${\Bbb Z}_{2}$  terminal singular points.
The physical Hodge numbers are obtained by
the Batyrev formula (\ref{batyrev}) or the Vafa formula 
of Landau-Ginzburg orbifold \cite{Vacua}: 
$(h^{1,1}, h^{1,3}, h^{1,2}, h^{2,2})=(1, 443, 0, 1820).$ \\
As $h^{1,1}=1$, we could say that the physical Hodge numbers are those of
the  original singular variety $X_{4}$.
Compare this with $X_{4}:={\Bbb P}_{(1,1,1,1,4,4)}[12]$
which has three ${\Bbb Z}_{4}$ singular points
which can be blown-up to ${\Bbb P}^{3}$ and 
$h^{1,1}=1+3$.\\
We conjecture that string theory 
on a Calabi-Yau fourfold $X_{4}$  which has at most terminal point 
singularities is well-defined at the perturbative level
as we can realize such models as Landau-Ginzburg
or torus orbifolds.
Hence we may  restrict ourselves here to the  Calabi-Yau fourfolds
the Newton polytopes of which are  reflexive. \\
Among them the models which can be described as  a Landau-Ginzburg orbifold
are  of importance  because 
(the mirror dual of) the intersection form:\\
$H^{2,2}\times H^{2,2} \rightarrow {\Bbb C}$ \cite{BB},
which contributes to the electric charge of F/M/IIA theory 
4/3/2-form gauge field $C_{(4/3/2)}$ \cite{BB,FMS,Mayr} according to
\begin{equation}
d \tilde{F}_{(7)} = -\frac{1}{2}F_{(4)}\wedge F_{(4)}
+I_{(8)}(R)+\delta^{(8)}(3/2/1\mbox{-Branes}),
\end{equation}
can be  easily computed  from the chiral ring 
${\cal R}:={\Bbb C}[x_{i}]/\partial_{i}W$ 
except for the absolutely few members that come from twisted sectors.

%
\newpage
\section{Smooth Models}
\subsection{Smooth Elliptic Calabi-Yau Fourfolds}
If the Ricci tensor $R_{i\overline{j}}(b)$ of a threefold $B_{3}$
is positive semi-definite, then  type IIB string on $B_{3}$
yields a N=1,
D=4 model which has generically no non-Abelian gauge symmetry,
as the elliptic Calabi-Yau fourfold $X_{4}$ over it is
smooth for a generic choice of the parameters.
These models have a virtue that we can easily compute the Euler number 
of $X_{4}$,
\begin{equation}
\frac{1}{24}\chi(X_{4})
=12 + 15 \ c_{1}(B_{3})\cdot c_{1}(B_{3})\cdot c_{1}(B_{3}),
\label{Euler}
\end{equation}
and the superpotential anomaly 
\begin{equation}
\chi(D_{3},{\cal O}_{D_{3}})
=-\frac{1}{2}\ W_{2}\cdot W_{2}\cdot c_{1}(B_{3}),\qquad W_{2}:=\pi(D_{3}),
\label{superpot3}
\end{equation}
where $D_{3}$ ($W_{2}$) is a Euclidean world volume of F/M/IIA
theory five-brane (type IIB theory three-brane)
which  induces a non-perturbative effect
\cite{BBS,Wi2}, and we have used 
the integration along the elliptic fiber \cite{SVW}.

A threefold is called Fano %
if its Ricci tensor is positive definite.
It serves as a good example of smooth Weierstrass models.\\
The toric Fano threefolds are completely classified \cite{Oda}
and are given  in Figure \ref{18Fanos}. They are called  as  
${\frak F}_{n}$, $1 \leq n \leq 18$.
\subsection{Phase Structure of ${\frak F}_{12}$}
\begin{figure}[h]
\psfig{file=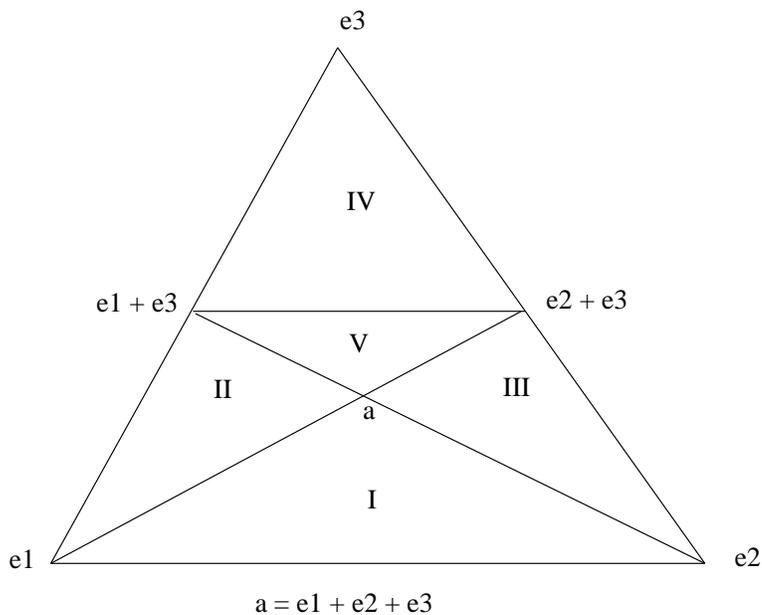,width=10cm}
\caption{Phase Diagram of ${\frak F}_{12}$ Model}
\label{F12 Model}
\end{figure}

Here we choose a specific example ${\frak F}_{12}$ to
explain blowing-up/down transitions
 shown in Figure \ref{18Fanos}. \\
The toric data of ${\frak F}_{12}$ is given by the following
homogeneous co-ordinates and the charge assignments;
\begin{alignat}{7}
 &{}         &\quad %
   &x_{1}   & \quad %
    &x_{2}  & \quad %
&x_{3} & \quad %
&x_{4} & \quad %
 &x_{5} & \quad %
 &x_{6}    \nonumber \\ %
&\lambda_{1} & \quad %
&0     & \quad %
&1     & \quad %
&1     & \quad %
&0     &\quad  %
  &0   & \quad %
&1     \nonumber \\ %
&\lambda_{2} & \quad %
 &1     & \quad %
 &1     & \quad %
&1     & \quad  %
&1     &   \quad %
 &0   & \quad %
 &0      \nonumber \\     %
&\lambda_{3} &\quad %
  &0     & \quad %
&0     & \quad %
&0     & \quad  %
&1     & \quad %
   &1   & \quad %
 &0.         %
\end{alignat}
The  total K\"ahler cone  ${\cal K}_{\mbox{tot}}$ 
is  spanned by the three vectors \\ 
${\cal K}_{\mbox{tot}}:={\Bbb R}_{\geq 0}
\langle e_{1},e_{2},e_{3} \rangle,$
which  is further subdivided into five K\"ahler cones 
according to the topology of the resulting toric threefold 
${\frak F}_{n}$, $n=1, 2, 3, 5, 12$.     \\ %
The phase diagram is given in Figure \ref{F12 Model}.
%
%
The  phase transition at each  phase boundary
is described as follows.
\begin{itemize}
\item  On the phase boundary of V and II,
the divisor   $\{ x_{6} = 0 \}$ $\cong$ ${\Bbb P}^{1}(x_2,x_3)\times
{\Bbb P}^{1}(x_1,x_4)$ of $ {\frak F}_{12}$  contracts to the two cycle
${\Bbb P}^{1}(x_2,x_3)$.
\item  On the boundary between V and III,
the divisor $\{ x_{5} = 0 \}$ $\cong$  ${\Bbb P}^{2}(x_1,x_2,x_3)$ 
of ${\frak F}_{12}$ contracts to a point.
\item On the phase boundary between V and IV,
the ${\Bbb P}^{1}$ fibers of
the divisor $\{x_{1} =0 \}$ $\cong$ ${\Bbb F}_{1}(x_2,x_3;x_5,x_6)$ 
of $ {\frak F}_{12} $ contracts.
\item On the phase boundary of III and  I,
the divisor  $\{ x_{6} = 0 \}$ $\cong $ ${\Bbb P}^{1}(x_1,x_4)
\times {\Bbb P}^{1}(x_2,x_3) $  of ${\frak F}_{5}$ 
contracts to ${\Bbb P}^{1}(x_1,x_4)$.
\item On the phase boundary between II and I,
the divisor $\{ x_{5} = 0 \}$ $\cong$ ${\Bbb P}^{2}(x_1,x_2,x_3)$  
of ${\frak F}_{3}$ contracts to a point. 
\end{itemize}
Indeed we expect these phase transitions%
\footnote{Although we don't take into account  any quantum corrections
to the K\"ahler moduli space,
we expect that pure classical treatment
here would suffice to see 
qualitative structures of the phase transitions.}
to occur  by changing the VEV of the
chiral supermultiplets associated with the cohomology class 
$H^{1,1}({\frak F}_{n})$.
%
%
%
\newpage
The phase structure of our model is summarized
 in Table \ref{B12-Kahler cones}.
\begin{table}[h]
\caption{Phase Structure of Toric Fano Threefolds}
\label{B12-Kahler cones}
\begin{tabular}{|l||l|l|l|}
\hline
{} & generators of K\"ahler  cone &  excluded set $F$ &  Fano threefold \\
\hline
{}    & $e_{1}$ & $\{x_{6}=0\}\cup $ & {}\\
 I    & $e_{2}$    & $\{x_{5}=0\}\cup $ & ${\frak F}_{1}$       \\
 {}    & $ e_{1}+e_{2}+e_{3}$
   & $\{x_{1}=x_{2}=x_{3}=x_{4}=0\}$ & { }   \\ 
\hline
{}    & $e_{1}$ & $\{x_{6}=0 \}\cup $ & {} \\
   II  & 
$ e_{1}+e_{3}$ & $\{x_{4}=x_{5}=0\}\cup $ & ${\frak F}_{3}$    \\
 {}    & $ e_{1}+e_{2}+e_{3}$
   & $\{x_{1}=x_{2}=x_{3}=0\}$ & { } \\ \hline
{}    & $e_{2}$ & $\{x_{5}=0\}\cup $ & {}\\
  III   &  $ e_{2}+e_{3} $  & $\{x_{1}=x_{4}=0\}\cup $ & ${\frak F}_{5}$    \\
    {} & $ e_{1}+e_{2}+e_{3}$
   & $\{x_{2}=x_{3}=x_{6}=0\}$ & { }\\ \hline
{}    & $ e_{3} $ & $\{ x_{1}=0 \}\cup $ & {}\\
 IV    & 
  $ e_{1}+e_{3} $ 
& $\{ x_{4}=x_{5}=0 \}\cup $ & $ {\frak F}_{2} $  \\
   {}  & $e_{2}+e_{3}$   & $\{ x_{2}=x_{3}=x_{6}=0 \}$ &  { } \\ \hline
{}    & $ e_{1}+e_{3}$
 & $\{ x_{1}=x_{4}=0 \}\cup $ & {}\\
V    &  $ e_{2}+e_{3}$  & $\{ x_{5}=x_{6}=0 \}
\cup $ & ${\frak F}_{12}$    \\
    {} & $ e_{1}+e_{2}+e_{3}$
   & $\{ x_{1}=x_{2}=x_{3}=0 \}$ & {}     \\        \hline
\end{tabular}
\end{table}
%
%

\subsection{Toric Fano Threefolds}
According to the diagram  of the eighteen 
toric Fano threefolds in Figure \ref{18Fanos},
there are three members that are on `the top
of the diagram':
${\frak F}_{18}$, ${\frak F}_{17}$ and
${\frak F}_{11}$.
The remaining members
 are obtained by the successive blowing-downs of 
one of the above three.

%
In Figure \ref{18Fanos}, a solid line with $n(l)$ 
means a blowing-down  of the exceptional
divisor ${\Bbb F}_{n}$ with the normal bundle of
type $(m,l)=(-1,l)$ (see section 8)
to ${\Bbb P}^{1}$, while
a dotted line means a blowing-down of the exceptional divisor
${\Bbb P}^{2}$ to a point.
Therefore it will suffice here to describe only the three top Fano threefolds. 
The remaining (non-trivial) members shall  be found  in 
the subsequent sections and Appendix A.
\newpage
%
\begin{figure}[h]
\setlength{\unitlength}{0.012000in}%
\begingroup\makeatletter
\def\x#1#2#3#4#5#6#7\relax{\def\x{#1#2#3#4#5#6}}%
\expandafter\x\fmtname xxxxxx\relax \def\y{splain}%
\ifx\x\y   
\gdef\SetFigFont#1#2#3{%
  \ifnum #1<17\tiny\else \ifnum #1<20\small\else
  \ifnum #1<24\normalsize\else \ifnum #1<29\large\else
  \ifnum #1<34\Large\else \ifnum #1<41\LARGE\else
     \huge\fi\fi\fi\fi\fi\fi
  \csname #3\endcsname}%
\else
\gdef\SetFigFont#1#2#3{\begingroup
  \count@#1\relax \ifnum 25<\count@\count@25\fi
  \def\x{\endgroup\@setsize\SetFigFont{#2pt}}%
  \expandafter\x
    \csname \romannumeral\the\count@ pt\expandafter\endcsname
    \csname @\romannumeral\the\count@ pt\endcsname
  \csname #3\endcsname}%
\fi
\endgroup
\begin{picture}(426,368)(683,422)
\thicklines
\put(928,668){\line( 2, 1){ 64.800}}
\put(1012,708){\vector( 3, 2){ 23.077}}
\put(785,706){\line( 3, 2){ 56.538}}
\put(859,754){\vector( 3, 2){ 48.923}}
\put(786,760){\line( 3,-2){ 63.462}}
\put(786,705){\line( 3,-1){ 53.400}}
\put(841,634){\vector( 3,-4){ 62.040}}
\put(786,641){\line( 3,-1){ 28.800}}
\put(786,641){\line( 4,-5){ 67.610}}
\put(865,544){\vector( 3,-4){ 38.520}}
\put(783,581){\line( 5, 1){ 15.961}}
\put(784,581){\line( 4,-5){ 63.122}}
\put(916,558){\line( 5, 6){ 41.393}}
\put(970,622){\vector( 3, 4){ 73.320}}
\put(859,489){\vector( 3,-4){ 43.920}}
\put(705,615){\line( 2,-1){ 14.800}}
\put(704,723){\line( 4,-5){ 26.244}}
\put(741,678){\vector( 3,-4){ 22.200}}
\put(705,724){\line( 5, 3){ 22.647}}
\put(745,748){\vector( 3, 2){ 19.846}}
\put(928,726){\line(2,-1){12}}
\put(956,713){\vector( 2,-1){ 86.800}}
\put(957,689){\vector( 1,-1){ 82}}
\put(704,723){\line( 4,-1){ 20}}
\put(787,703){\line( 4,-5){ 43.512}}
\put(922,602){\line( 3,-1){ 57.900}}
\put(1001,577){\vector( 3,-1){ 39.600}}
\put(934,778){\line( 1,-2){ 50.200}}
\put(989,669){\vector( 1,-2){ 51.400}}
\put(822,588){\vector( 4, 1){ 72.941}}
\put(835,626){\vector( 3,-1){ 59.400}}
\multiput(934,778)(22.87012,-11.43506){5}{\line( 2,-1){ 12.920}}
\put(1038,726){\vector( 2,-1){0}}
\put(786,705){\line( 5,-4){ 58.293}}
\put(859,646){\vector( 4,-3){ 40.320}}
\put(936,777){\line( 2,-3){ 82.923}}
\put(1030,640){\vector( 2,-3){ 16.462}}
\put(736,600){\vector( 2,-1){ 27.600}}
\put(746,713){\vector( 4,-1){ 19.059}}
\put(970,642){\vector( 2,-1){ 65.200}}
\put(928,719){\line( 1,-1){ 14}}
\put(871,702){\vector( 3,-2){ 35.308}}
\multiput(1055,613)(13.84205,20.76308){4}{\line( 2, 3){  8.012}}
\put(1105,687){\vector( 2, 3){0}}
\put(1056,719){\line(2,-1){12}}
\put(864,682){\vector( 3,-1){ 35.700}}
\put(1083,707){\vector( 2,-1){ 15.600}}
\put(929,662){\line( 2,-1){ 20.735}}
\put(1065,707){\makebox(0,0)[lb]{\smash{\SetFigFont{8}{9.6}{ }0(1)}}}
\put(952,645){\makebox(0,0)[lb]{\smash{\SetFigFont{8}{9.6}{ }0(0)}}}
\put(989,702){\makebox(0,0)[lb]{\smash{\SetFigFont{8}{9.6}{ }1(-1)}}}
\put(843,745){\makebox(0,0)[lb]{\smash{\SetFigFont{8}{9.6}{ }0(0)}}}
\put(843,683){\makebox(0,0)[lb]{\smash{\SetFigFont{8}{9.6}{ }0(1)}}}
\put(827,639){\makebox(0,0)[lb]{\smash{\SetFigFont{8}{9.6}{ }1(-1)}}}
\put(813,627){\makebox(0,0)[lb]{\smash{\SetFigFont{8}{9.6}{ }1(-1)}}}
\put(852,547){\makebox(0,0)[lb]{\smash{\SetFigFont{8}{9.6}{ }1(0)}}}
\put(801,583){\makebox(0,0)[lb]{\smash{\SetFigFont{8}{9.6}{ }0(0)}}}
\put(846,493){\makebox(0,0)[lb]{\smash{\SetFigFont{8}{9.6}{ }0(0)}}}
\put(959,612){\makebox(0,0)[lb]{\smash{\SetFigFont{8}{9.6}{ }0(1)}}}
\put(720,602){\makebox(0,0)[lb]{\smash{\SetFigFont{8}{9.6}{ }0(0)}}}
\put(728,682){\makebox(0,0)[lb]{\smash{\SetFigFont{8}{9.6}{ }1(0)}}}
\put(729,740){\makebox(0,0)[lb]{\smash{\SetFigFont{8}{9.6}{ }0(1)}}}
\put(940,714){\makebox(0,0)[lb]{\smash{\SetFigFont{8}{9.6}{ }1(1)}}}
\put(846,649){\makebox(0,0)[lb]{\smash{\SetFigFont{8}{9.6}{ }1(0)}}}
\put(981,576){\makebox(0,0)[lb]{\smash{\SetFigFont{8}{9.6}{ }0(0)}}}
\put(982,671){\makebox(0,0)[lb]{\smash{\SetFigFont{8}{9.6}{ }1(0)}}}
\put(1019,645){\makebox(0,0)[lb]{\smash{\SetFigFont{8}{9.6}{ }0(1)}}}
\put(723,713){\makebox(0,0)[lb]{\smash{\SetFigFont{8}{9.6}{ }1(-1)}}}
\put(941,695){\makebox(0,0)[lb]{\smash{\SetFigFont{8}{9.6}{ }2(-1)}}}
\put(850,707){\makebox(0,0)[lb]{\smash{\SetFigFont{8}{9.6}{ }1(-1)}}}
\put(1045,560){\makebox(0,0)[lb]{${\frak F}_{2}$}}
%
%
\put(912,779){\makebox(0,0)[lb]{${\frak F}_{12}$}}
\put(1042,722){\makebox(0,0)[lb]{${\frak F}_{5}$}}
%
%
\put(1047,667){\makebox(0,0)[lb]{${\frak F}_{4}$}}
\put(903,604){\makebox(0,0)[lb]{${\frak F}_{9}$}}
%
\put(905,424){\makebox(0,0)[lb]{${\frak F}_{6}$}}
%
\put(683,722){\makebox(0,0)[lb]{${\frak F}_{18}$}}
%
\put(688,617){\makebox(0,0)[lb]{${\frak F}_{17}$}}
%
\put(908,725){\makebox(0,0)[lb]{${\frak F}_{11}$}}
%
%
\put(765,640){\makebox(0,0)[lb]{${\frak F}_{14}$}}
%
\put(765,581){\makebox(0,0)[lb]{${\frak F}_{13}$}}
\put(1042,601){\makebox(0,0)[lb]{${\frak F}_{3}$}}
%
%
\put(766,705){\makebox(0,0)[lb]{${\frak F}_{15}$}}
\put(908,664){\makebox(0,0)[lb]{${\frak F}_{10}$}}
%
\put(1101,695){\makebox(0,0)[lb]{${\frak F}_{1}$}}
%
\put(766,760){\makebox(0,0)[lb]{${\frak F}_{16}$}}
%
\put(905,544){\makebox(0,0)[lb]{${\frak F}_{8}$}}
%
\put(903,482){\makebox(0,0)[lb]{${\frak F}_{7}$}}
%
\end{picture}
\caption{Toric Fano Threefolds}
\label{18Fanos}
\end{figure}
\newpage

The toric data of ${\frak F}_{18}$ is given by %
the following charge assignment.
\begin{alignat}{8}
&x_{7} & \quad %
&x_{8} & \quad %
&x_{1} & \quad %
&x_{3} & \quad %
&x_{5} & \quad %
&x_{2} & \quad %
&x_{4} & \quad %
&x_{6}  \nonumber \\
&1 & \quad %
&1 & \quad %
-&1 & \quad %
&0 & \quad %
&0 & \quad %
&0 & \quad %
&0 & \quad %
&0   \nonumber \\
&0 & \quad %
&0 & \quad %
&1 & \quad %
&1 & \quad %
&0 & \quad %
-&1 & \quad %
&0 & \quad %
&0 \nonumber \\ %
&0 & \quad %
&0 & \quad %
&0 & \quad %
&1 & \quad %
&1 & \quad %
&0 & \quad %
-&1 & \quad %
&0  \nonumber \\ %
&0 & \quad %
&0 & \quad %
&1 & \quad %
&0 & \quad %
&1 & \quad %
&0 & \quad %
&0 & \quad %
-&1 \nonumber \\ %
&0 & \quad %
&0 & \quad %
-&1 & \quad %
-&1 & \quad %
-&1 & \quad %
&1 & \quad %
&1 & \quad %
&1.         \label{F18}%
\end{alignat}
We have chosen the combinations of charges 
so that the excluded set is 
manifest in this expression, namely
the coordinates that has a positive charge, e.g. 
$(x_{7},x_{8})$, $(x_{1},x_{3})$, $(x_{2},x_{4},x_{6})$, %
cannot  simultaneously be zero.\\
It is clear in (\ref{F18}) that ${\frak F}_{18}$ has a structure of 
non-trivial ${\Bbb S}_{3}$ bundle over ${\Bbb P}^{1}$,
where ${\Bbb S}_{d}$ is the del Pezzo surface of degree $d$ \cite{MV2}.\\
The blowing-downs of the divisors $\{ x_{2,6}=0\}$, %
$\{ x_{4}=0\}$, and $\{ x_{3,5}=0\}$ give rise to
${\frak F}_{15}$, ${\frak F}_{16}$, and ${\frak F}_{14}$
respectively.\\ 
%
${\frak F}_{17}$ is isomorphic to 
${\Bbb S}_{3}\times {\Bbb P}^{1}$.
Its toric data is as follows;
\begin{alignat}{8}
&x_{7} & \quad %
&x_{8} & \quad %
&x_{1} & \quad %
&x_{3} & \quad %
&x_{5} & \quad %
&x_{2} & \quad %
&x_{4} & \quad %
&x_{6}  \nonumber \\
&1 & \quad %
&1 & \quad %
&0 & \quad %
&0 & \quad %
&0 & \quad %
&0 & \quad %
&0 & \quad %
&0   \nonumber \\
&0 & \quad %
&0 & \quad %
&1 & \quad %
&1 & \quad %
&0 & \quad %
-&1 & \quad %
&0 & \quad %
&0 \nonumber \\ %
&0 & \quad %
&0 & \quad %
&0 & \quad %
&1 & \quad %
&1 & \quad %
&0 & \quad %
-&1 & \quad %
&0  \nonumber \\ %
&0 & \quad %
&0 & \quad %
&1 & \quad %
&0 & \quad %
&1 & \quad %
&0 & \quad %
&0 & \quad %
-&1 \nonumber \\ %
&0 & \quad %
&0 & \quad %
-&1 & \quad %
-&1 & \quad %
-&1 & \quad %
&1 & \quad %
&1 & \quad %
&1         \label{F17}%
\end{alignat}
Any blowing-down of  $\{ x_{i}=0 \}$, %
$1\leq i \leq 6 $ %
leads to ${\frak F}_{13}$. \\
The toric data of ${\frak F}_{11}$ is as follows.
\begin{alignat}{6}
&x_{1} & \quad %
&x_{2} & \quad %
&x_{3} & \quad %
&x_{4} & \quad %
&x_{5} & \quad %
&x_{6} \nonumber \\ %
&0   & \quad %
&1   & \quad %
&1   & \quad %
&0   & \quad %
-&2   & \quad %
&1      \nonumber \\ %
-&1   & \quad %
&0   & \quad %
&0   & \quad %
&1   & \quad %
&0   & \quad %
&1      \nonumber \\
&1   & \quad %
&0   & \quad %
&0   & \quad %
&0   & \quad %
&1   & \quad %
-&1.          \label{F11}%
\end{alignat}
There are two exceptional divisors:
the blowing-down of $\{ x_{6}=0 \}$  
and that of $\{ x_{1}=0 \}$
give ${\frak F}_{3}$ and ${\frak F}_{4}$ respectively. 
%
%
\subsection{Quantum Numbers of Toric Fano Models}
The Euler number of the toric Fano model can be computed
by intersection paring formula (\ref{Euler}).
The Hodge numbers of it can also be known by analyzing
the combinatoric structure of 
the Newton polytope $\Delta$ and the polar polytope $\Delta^{*}$
(\ref{batyrev}). \\
We collect in  Table \ref{ToricFano} the quantum numbers 
of the toric Fano threefold models.
\begin{table}[h]
\begin{center}
\caption{Quantum  Numbers of Toric Fano  Models}
\label{ToricFano}
\begin{tabular}{|l||l|l|l|l|l|}\hline %
Model&  %
$h^{1,1}$ & $h^{1,3}$ & %
$h^{1,2}$ & $h^{2,2}$ & $ \chi(X_{4})/24$ \\  \hline %
${\frak F}_{1}$ & 2 & 3878 & 0 & 15564 &  972 \\  \hline
${\frak F}_{2}$ & 3 & 3277  & 0 & 13164 & 822 \\  \hline
${\frak F}_{3}$ & 3  & 3397 & 0 & 13644  & 852  \\  \hline
${\frak F}_{4}$ & 3  & 3757 & 0 & 15084 & 942 \\  \hline
${\frak F}_{5}$ & 3  & 3277 & 0 & 13164 &  822 \\  \hline
${\frak F}_{6}$ & 4 &  2916 & 0 & 11724 & 732 \\  \hline
${\frak F}_{7}$  & 4 & 3156 & 0 & 12684  & 792\\  \hline
${\frak F}_{8}$ & 4 & 2676 & 0 & 10764  & 672 \\  \hline
${\frak F}_{9}$  & 4 & 2916 & 0 & 11724 & 732\\  \hline
${\frak F}_{10}$  & 4 & 3036 & 0 & 12204 & 762\\  \hline
${\frak F}_{11}$  & 4 & 3036  & 0 & 12204 & 762\\  \hline
${\frak F}_{12}$  & 4 & 2796 & 0 & 11244 & 702\\  \hline
${\frak F}_{13}$  & 5 & 2555 & 0 & 10284 & 642\\  \hline
${\frak F}_{14}$  & 5 & 2675 & 0 & 11124 & 672\\  \hline
${\frak F}_{15}$  & 5 & 2435 & 0 & 9804 & 612\\  \hline
${\frak F}_{16}$  & 5 & 2795 & 0 & 11244 & 702\\  \hline
${\frak F}_{17}$  & 6 & 2194 & 0 & 8844 & 552\\  \hline
${\frak F}_{18}$  & 6 & 2194 & 0 & 8844 & 552\\  \hline
\end{tabular}
\end{center}
\end{table}
We also give some identifications of the above models.\\
${\frak F}_{1}$ $= $  ${\Bbb P}^{3}$, \ %
${\frak F}_{2}$ $=$ ${\Bbb P}^{2}\times {\Bbb P}^{1}$, %
\ %
${\frak F}_{6}$ $= $  ${\Bbb P}^{1}$ $\times$ ${\Bbb P}^{1}$ $\times$ 
${\Bbb P}^{1}$, \ %
${\frak F}_{9}$ $=$ ${\Bbb F}_{1}$ $\times$ ${\Bbb P}^{1}$,
\ %
${\frak F}_{13}$ $=$ ${\Bbb S}_{2}$ $\times$ ${\Bbb P}^{1}$.
%
%

%
\newpage
\subsection{Other Smooth Models}
\begin{figure}[h]
\psfig{file=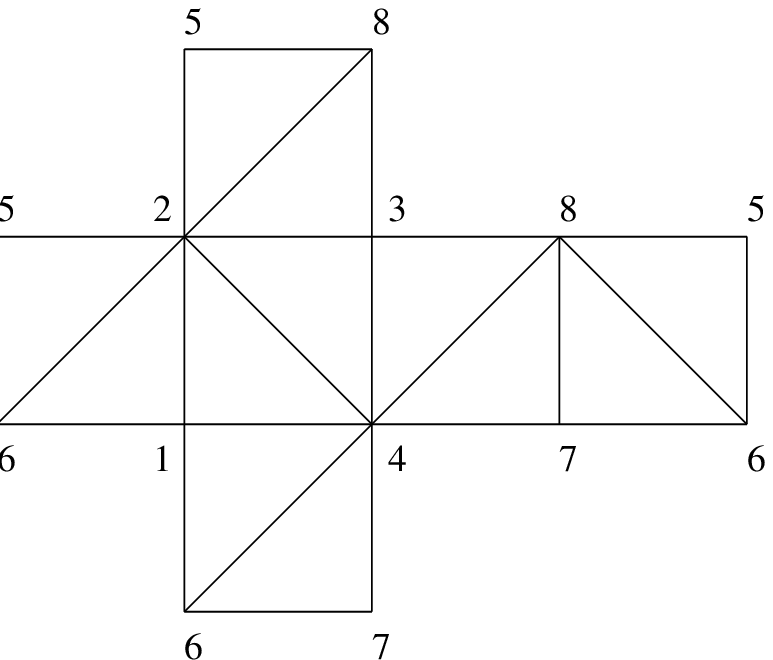,width=8cm}
\caption{Fan of ${\cal  T}$}  
\label{P3-blowup}
\end{figure}

A Ricci semi-positive toric threefold which is not Fano
has a curve $C$ $\cong$ ${\Bbb P}^{1}$, called 
$(-2)$ curve, about  which the type IIB axion-dilaton $\tau$
is constant, %
i.e. the normal bundle ${\cal N}$
of $C$ is a non-compact Calabi-Yau threefold,
which has the following expression:
\begin{equation}
{\cal N}={\cal O}_{{\Bbb P}^{1}}(a)\oplus {\cal O}_{{\Bbb P}^{1}}(b), \quad %
a+b=-2.
\end{equation}
This is in contrast to  the case of a Fano threefold,
where {\it any}  ${\Bbb P}^{1}$ in it has a normal bundle %
of the type $(a,b)$ with
$ a+b$  $\geq -1$, 
pierced by $10(a+b)+18$ seven-branes \cite{Wi1}.\\
The existence of a $(-2)$ curve  $C$ 
in Ricci semi-positive threefolds is of physical importance because
a string with N=2 supersymmetry is produced by
wrapping type IIB three-brane around $C$. %
We give some examples of
Ricci semi-positive models and their
$(-2)$ curves.\\
First there are models which has ${\Bbb P}^{1}$ of type
$(a,b)=(-1,-1)$.
It is well-known that such  ${\Bbb P}^{1}$
induces a  flop transition.
For example, 
each of the following divisor: $\{ x_{1}=0 \}$ of ${\frak F}_{18}$,
$\{ x_{1}=0 \}$ of ${\frak F}_{14}$, or  
$\{ x_{6}=0 \}$ of ${\frak F}_{7}$ 
is isomorphic to ${\Bbb F}_{0}$ and 
can be blown down to give a  Ricci semi-positive threefold 
with a $(-1,-1)$ curve.\\
%
For another example we present ${\cal T}$ which is obtained by
blowing-up \\
${\Bbb P}^{3}(x_{2},x_{4},x_{6},x_{8})$ %
around  the four points %
$(1,0,0,0)$, $(0,1,0,0)$, $(0,0,1,0)$ and $(0,0,0,1)$ \cite{Fulton}. %
The toric data of ${\cal T}$ is as follows.
\begin{alignat}{8}
&x_{2} & \quad %
&x_{4} & \quad %
&x_{6} & \quad %
&x_{8} & \quad %
&x_{1} & \quad %
&x_{3} & \quad %
&x_{5} & \quad %
&x_{7} \nonumber  \\ %
-&2   & \quad %
-&2   & \quad %
-&2   & \quad %
-&2   & \quad %
&1   & \quad %
&1   & \quad %
&1  & \quad %
&1   \nonumber  \\ %
&1   & \quad %
&1   & \quad %
&1   & \quad %
&0   & \quad %
-&1   & \quad %
&0   & \quad %
&0   & \quad %
&0  \nonumber \\ %
&1   & \quad %
&1   & \quad %
&0   & \quad %
&1   & \quad %
&0   & \quad %
-&1   & \quad %
&0   & \quad %
&0   \nonumber \\ %
&1   & \quad %
&0   & \quad %
&1   & \quad %
&1   & \quad %
&0   & \quad %
&0   & \quad %
-&1   & \quad %
&0   \nonumber \\ %
&0   & \quad %
&1   & \quad %
&1   & \quad %
&1   & \quad %
&0   & \quad %
&0   & \quad %
&0   & \quad %
-&1.           %
\end{alignat}
Here we give  the quantum numbers of the ${\cal T}$ model
computed from Batyrev formula:
\begin{equation}
(h^{1,1}, h^{1,3}, h^{1,2}, h^{2,2}) = (6, 1954, 0, 7884), \quad %
\chi(X_{4})/24=492.  
\end{equation}
To see the flops that ${\cal  T}$ admits,
it is useful to show  the triangulation of ${\Bbb S}^{2}$ 
defined by the fan of ${\cal  T}$  \cite{Oda} in Figure \ref{P3-blowup}.
\newpage
\begin{figure}[h]
\psfig{file=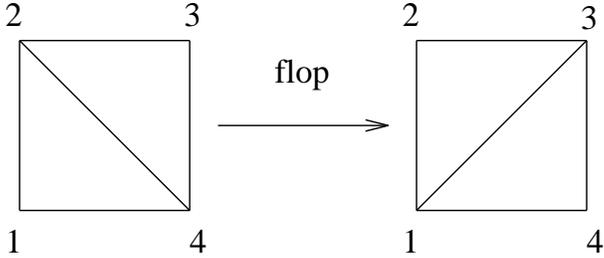,width=8cm}
\caption{Flop Transition}
\label{flop-picture}
\end{figure} 

%
The flop transition shown in Figure \ref{flop-picture} means that
${\Bbb P}^{1}(x_{1},x_{3})$ \\
$\cong$ $\{ x_{2}=x_{4}=0 \}$
is contracted and replaced by ${\Bbb P}^{1}(x_{2},x_{4})$
$\cong$ $\{ x_{1}=x_{3}=0 \}$.  \\
From the fan in Figure \ref{P3-blowup}, %
we see that ${\cal T}$ admits six flops;
\begin{eqnarray}
\ {\Bbb P}^{1}(x_{3},x_{5})
\longleftrightarrow {\Bbb P}^{1}(x_{2},x_{8}),
&\ & %
 \ {\Bbb P}^{1}(x_{1},x_{5})
\longleftrightarrow {\Bbb P}^{1}(x_{2},x_{6}), \nonumber \\
 \ {\Bbb P}^{1}(x_{1},x_{7})
\longleftrightarrow {\Bbb P}^{1}(x_{4},x_{6}),
 &\  & %
  \ {\Bbb P}^{1}(x_{1},x_{3})
\longleftrightarrow {\Bbb P}^{1}(x_{2},x_{4}), \\
 \ {\Bbb P}^{1}(x_{3},x_{7})
\longleftrightarrow {\Bbb P}^{1}(x_{4},x_{8}),
&\ &%
 \ {\Bbb P}^{1}(x_{5},x_{7})
\longleftrightarrow {\Bbb P}^{1}(x_{6},x_{8}).\nonumber
\end{eqnarray}
Accordingly, ${\cal T}$ model gives rise to six tensionless strings
carrying $N$ $=$ $2$ supersymmetry. \\
%
%
Second  ${\Bbb O}_{2}$ model is defined by the following toric data:
\begin{alignat}{6}
&{} & \quad %
&x_{1} & \quad %
&x_{2} & \quad %
&y_{1} & \quad %
&y_{2} & \quad %
&y_{3} \nonumber \\  
&{} & \quad %
&1  & \quad %
&1  & \quad %
&2  & \quad %
&2  & \quad %
&0    \nonumber \\
&{}  & \quad %
&0  & \quad %
&0  & \quad %
&1  & \quad %
&1  & \quad %
&1,          %
\end{alignat}
where the excluded set is 
$\{ x_{1}=x_{2}=0 \}$ %
$\cup$ %
$\{ y_{1}=y_{2}=y_{3}=0\}$. \\
The curve $\{ y_{3}=x_{i}=0 \}$ has the normal bundle 
${\cal O}_{{\Bbb P}^{1}}(-2)\oplus {\cal O}_{{\Bbb P}^{1}}$.\\
Third ${\Bbb G}_{3}$ model is defined by the toric data:
\begin{alignat}{6}
&{} & \quad %
&x_{1} & \quad %
&x_{2} & \quad %
&x_{1} & \quad %
&y_{1} & \quad %
&y_{2} \nonumber \\  
&{} & \quad %
&1  & \quad %
&1  & \quad %
&1  & \quad %
&3  & \quad %
&0    \nonumber \\
&{}  & \quad %
&0  & \quad %
&0  & \quad %
&0  & \quad %
&1  & \quad %
&1,          %
\end{alignat}
where the excluded set is 
$\{ x_{1}=x_{2}=x_{3}=0 \}$ %
$\cup$ %
$\{ y_{1}=y_{2}=0\}$. \\
The curve defined by $\{y_{2}=x_{i}=0\}$
has the normal bundle 
${\cal O}_{{\Bbb P}^{1}}(-3)\oplus {\cal O}_{{\Bbb P}^{1}}(1)$. 
%

%
%
%
\newpage
\section{${\Bbb P}^{1}$ bundle over ${\Bbb P}^{2}$ models}
\subsection{Model}
Here we treat the threefold which we call ${\Bbb G}_{n}$ 
 defined by the toric data below;
%
\begin{alignat}{6}
 &{}         &\quad   &x_{1}   & \quad     &x_{2}  & \quad 
&x_{3} & \quad &y_{1} & \quad  &y_{2}     \nonumber \\ %
&\lambda & \quad &1     & \quad &1     & \quad &1     & \quad 
&n     &\quad    &0       \nonumber \\ %
&\mu &\quad   &0     & \quad &0     & \quad 
&0     & \quad  &1     & \quad    &1            
\end{alignat}
%
%
%
%
%
%
\begin{equation}
{\Bbb G}_{n}:=
 \{ {\Bbb C}^{5}-F \}/{\Bbb C}^{*}_{\lambda}\times {\Bbb C}^{*}_{\mu},
\end{equation}
where the excluded set is  %
$E=\{ x_{1}=x_{2}=x_{3}=0 \}\cup \{y_{1}=y_{2}=0\}$.\\
It has  a ${\Bbb P}^{1}$ fiber bundle structure over ${\Bbb P}^{2}$,
$\pi: {\Bbb G}_{n}(x,y)\rightarrow {\Bbb P}^{2}(x)$.
Thus type IIB compactification on ${\Bbb G}_{n}$
is dual to heterotic compactification on   %
the elliptic Calabi-Yau threefold $Y_{3}$
 over ${\Bbb P}^{2}$, which can also  be  realized by the weighted projective 
hypersurface ${\Bbb P}_{(1,1,1,6,9)}[18]$.\\
${\Bbb G}_{n}$ model in D=4 is close analogue of
${\Bbb F}_{n}$ model in D=6 \cite{MV1,MV2}.
The toric data for  the elliptic 
Calabi-Yau fourfold over ${\Bbb G}_{n}$
is given by
\begin{alignat}{9}
 &{}         &\quad %
 &x_{1}   & \quad %
 &x_{2}  & \quad  %
&x_{3} &  \quad   %
&y_{1} & \quad   %
&y_{2} & \quad   %
&z_{1} & \quad  %
&z_{2} & \quad  %
&z_{3}    \nonumber \\ %
&\lambda & \quad  %
&1 & \quad %
&1    & \quad  %
 &1  & \quad   %
&n     &\quad  %
&0   & \quad   %
&0  & \quad    %
 2(n&+3)  & \quad %
 3(n&+3)     \nonumber \\ %
&\mu &\quad  %
  &0     & \quad     %
&0     & \quad   %
&0  & \quad      %
&1     & \quad   %
&1     & \quad   %
&0   & \quad     %
&4    & \quad    %
&6         \nonumber \\ %
&\nu & \quad  %
&0     & \quad     %
&0     & \quad     %
&0 & \quad         %
&0     & \quad     %
&0     &   \quad   %
 &1   & \quad      %
&2  & \quad        %
&3                 %
\end{alignat}
%
%
%
%
It can be seen that for $0 \leq n\leq 3$,
${\Bbb G}_{n}$ model has generically no non-Abelian gauge symmetries.
The first three of them are identified with the toric Fano threefolds
encountered before;
${\Bbb G}_{0}\cong {\frak F}_{2}$, %
${\Bbb G}_{1}\cong {\frak F}_{3}$, %
${\Bbb G}_{2}\cong {\frak F}_{4}$. \\ %
For $4\leq n $,  several seven-branes 
inevitably coincide on the exceptional
cross section divisor $\{ y_{2}= 0\}$, producing  
non-Abelian gauge symmetry. \\
 We have checked using the program PORTA \cite{Porta}
 that for $n\geq 3$  the Newton polytope  of the elliptic 
Calabi-Yau fourfold is isomorphic to that of \\
 ${\Bbb P}_{(1,1,1,n,2(n+3),3(n+3))}[6(n+3)],$ 
which is reflexive for $ n\leq 18 $.\\
Thus we can analyze the  spectrum by going to D=3 M theory
Coulomb phase \cite{MV1} and then using the Vafa formula 
for the Landau-Ginzburg %
orbifold \cite{Vacua} if 
${\Bbb P}_{(1,1,1,n,2(n+3),3(n+3))}[6(n+3)]$ %
admits a transverse hypersurface \cite{CLS},
which occurs for $n=3,4,6,9,12,17,18$. %
We can also apply the Batyrev formula (\ref{batyrev})
even for the remaining cases \cite{COK}
as in the case of D=6 \cite{CF}.
\newpage
We list the quantum numbers of the models in Table \ref{Gn-models}.
\begin{table}[h]
\begin{center}
\caption{Quantum Numbers of ${\Bbb G}_{n}$ Models}
\label{Gn-models}
\begin{tabular}{|l||l|l|l|l|l|c|}\hline
{}& $h^{1,1}$ &$h^{1,3}$ &$h^{1,2}$ &$h^{2,2}$ &
$\chi(X_{4})/24$ & singularity type\\  \hline
${\Bbb G_{0}}$ & 3 & 3277 & 0 & 13164 & 822 & - \\  \hline
${\Bbb G_{1}}$ & 3 & 3397 & 0 & 13644 & 852 & - \\  \hline
${\Bbb G_{2}}$ & 3 & 3757 & 0 & 15084 & 942 & - \\  \hline
${\Bbb G_{3}}$ & 3 & 4358 & 1 & 17486 & 1092 & - \\  \hline
${\Bbb G_{4}}$ & 4 & 5187 & 0 & 20808 & 1299+3/4 & $III$ \\  \hline
${\Bbb G_{5}}$ & 5 & 6191 & 0 & 24828 & 1551 & ${I_{0}^{*ns}}$ \\  \hline
${\Bbb G_{6}}$ & 7 & 7341 & 0 & 29436 & 1839 & $I_{0}^{*s}$ \\  \hline
${\Bbb G_{7}}$ & 7 & 8957 & 0 & 35900 & 2243 & ${IV^{*ns}}$ \\  \hline
${\Bbb G_{8}}$ & 7 & 10045 & 0 & 40252 & 2515 & ${IV^{*ns}}$ \\  \hline
${\Bbb G_{9}}$ & 9 & 11587 & 0 & 46428 & 2901 & $IV^{*s}$ \\  \hline
${\Bbb G_{10}}$ & 10 & 13255 & 0 & 53104 & 3318+1/4 & ${III^{*}}$ \\  \hline
${\Bbb G_{11}}$ & 10 & 15046 & 0 & 60268 & 3766 & ${III^{*}}$ \\  \hline
${\Bbb G_{12}}$ & 10 & 16959 & 0 & 67920 & 4244+1/4 &
$III^{*}$ \\  \hline
${\Bbb G_{13}}$ & 12 & 18994 & 6 & 76056 & 4752 & ${II^{*}}$ \\  \hline
${\Bbb G_{14}}$ & 12 & 21151 & 3 & 84690& 5292 & ${II^{*}}$ \\  \hline
${\Bbb G_{15}}$ & 12 & 23429 & 1 & 93806 & 5862 & ${II^{*}}$ \\  \hline
${\Bbb G_{16}}$ & 12 & 25828 & 0 & 103404 & 6462 & ${II^{*}}$ \\  \hline
${\Bbb G_{17}}$ & 12 & 28348 & 0 & 113484 & 7092 &
$II^{*}$  \\  \hline
${\Bbb G_{18}}$ & 11 & 30989 & 0 & 124044 & 7752 &
$II^{*}$ \\  \hline
\end{tabular}
\end{center}
\end{table}
The singularity type 
in  Table \ref{Gn-models} means the one at $\{ y_{2}=0\}$.\\
As for the non-Abelian gauge symmetry 
localized at  $\{ y_{2}=0\}$,
 we remark that the prediction (\ref{vector})
of the rank of the gauge group $(h^{1,1}-3)$ seems consistent with
the singularity type of the elliptic fibration
analyzed by \cite{MV2,AG,sixmen}:
$\mbox{SU}_{2}\ (n=4)$, %
$\mbox{G}_{2} \ (n=5),$
$\mbox{SO}_{8}\ (n=6)$, %
$\mbox{F}_{4} \ (n=7,8),$ %
$\mbox{E}_{6}\ (n=9)$,  %
$\mbox{E}_{7}\ (n=10,11,12)$, %
$\mbox{E}_{8}\ (n=13-18)$. %
The reduction of the gauge groups to   non-simply laced ones 
at the singularities of non-split type:
$\mbox{SU}_{2n}\rightarrow \mbox{Sp}_{n}$,
$\mbox{SO}_{8}\rightarrow \mbox{G}_{2}$,
$\mbox{E}_{6} \rightarrow \mbox{F}_{4}$
have been explained in \cite{AG,sixmen}.

\newpage
\subsection{${\Bbb G}_{n}/{\Bbb G}_{n+1}$ Transition}
\begin{figure}[h]
\psfig{file=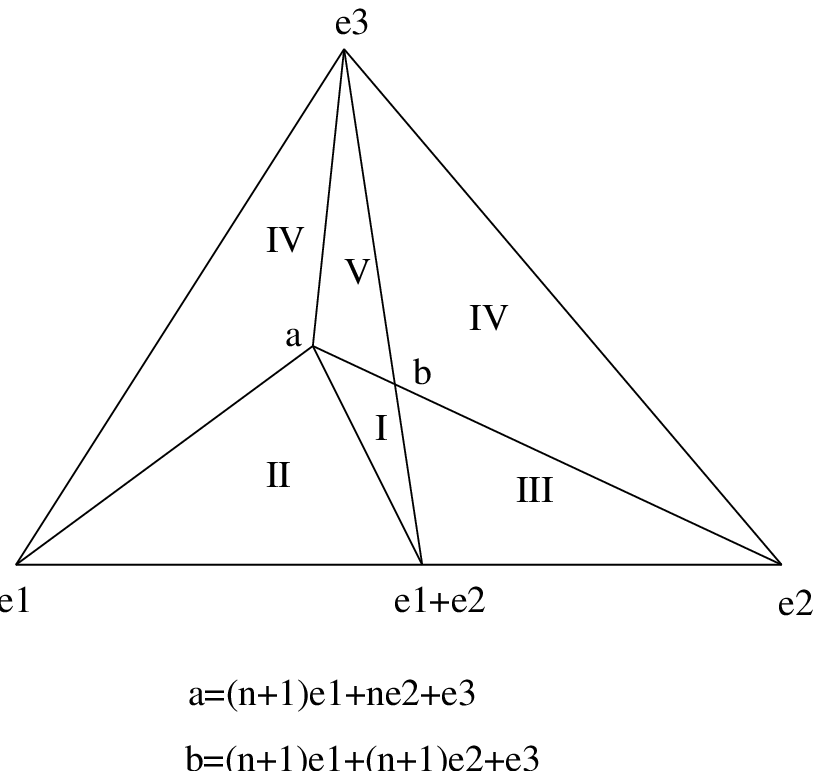,width=9cm}
\caption{ Phase Diagram of ${\Bbb H}_{n}$ Model}
\label{transition}
\end{figure}

If we blow up ${\Bbb G}_{n+1}$
around  the toric 2-cycle  $\{ y_{1}=x_{1}=0\}$ $\cong$ ${\Bbb P}^{1}$,
we obtain the toric threefold ${\Bbb H}_{n}$
associated with the toric data;
%
\begin{alignat}{7}
&{} & \quad
&x_{1} & \quad %
&x_{2} & \quad %
&x_{3} & \quad %
&x_{4} & \quad %
&x_{5} & \quad %
&x_{6}       \nonumber \\
&\lambda_{1} & \quad %
&1    & \quad %
&0    & \quad %
&1    & \quad %
&1    & \quad %
n&+1    & \quad %
&0  \nonumber \\
&\lambda_{2} & \quad %
&0    & \quad %
&1    & \quad %
&1    & \quad %
&1    & \quad %
&n    & \quad %
&0       \nonumber \\ %
&\lambda_{3}    & \quad %
&0    & \quad %
&0    & \quad %
&0    & \quad %
&0    & \quad %
&1    & \quad %
&1           %
\end{alignat}
%
%
%
%
%
%
 with the excluded set 
\begin{equation}
E=\{ x_1=x_5=0 \}\cup \{x_2=x_6=0\} \cup \{ x_1=x_3=x_4=0 \}.
\end{equation}
There is an identification; %
%
${\Bbb H}_{1}$ $\cong$ ${\frak F}_{11}$.
%
As ${\Bbb H}_{n}$ can also be obtained by blowing
${\Bbb G}_{n}$ along the curve $\{ x_{1}=y_{2}=0 \}$
on the exceptional cross section,
we see that ${\Bbb H}_{n}$ model interpolates between two vacua
${\Bbb G}_{n}$ and ${\Bbb G}_{n+1}$
via blowing-up/down of exceptional divisors. 
The total K\"ahler cone is generated by $ e_{1},e_{2},e_{3} $
and subdivided into six K\"ahler cones as shown in Figure \ref{transition}. 

%
\newpage
The phase structure is summerized in Table \ref{Hn-Kahler cones}.
\begin{table}[h]
\caption{Phase Structure of ${\Bbb H}_{n}$}
\label{Hn-Kahler cones}
\begin{flushleft}
\begin{tabular}{|l||l|l|l|}
\hline
{} & generators of K\"ahler cone &  excluded set $E$ & toric threefold \\
\hline
%
{}    & $ e_{1}+e_{2}$, 
& $\{x_{2}=x_{6}=0\}\cup $ & {}\\
 I    & 
$(n+1)e_{1}+ne_{2}+e_{3}$,
   & $\{x_{1}=x_{5}=0\}\cup $ & ${\Bbb H}_{n}$       \\
 {}    & $ (n+1)e_{1}+(n+1)e_{2}+e_{3}$  
 & $\{x_{1}=x_{3}=x_{4}=0\}$ & {}   \\ 
\hline
%
{}    & $ e_{1}$,
& $\{x_{1}=0 \}\cup $ & {} \\
   II  & 
$e_{1}+e_{2}$,    
& $\{x_{5}=x_{6}=0\}\cup $ & ${\Bbb G}_{n}$                          \\
 {}    & $(n+1)e_{1}+ne_{2}+e_{3}  $ 
  & $\{x_{2}=x_{3}=x_{4}=0\}$ & {} \\ \hline
%
{}    & $e_{2}$, 
& $\{x_{2}=0\}\cup $ & {}\\
  III   &
 $e_{1}+e_{2}$,    
& $\{x_{5}=x_{6}=0\}\cup $ & ${\Bbb G}_{n+1}$ \\
    {} &  $(n+1)e_{1}+(n+1)e_{2}+e_{3}$ 
  & $\{x_{1}=x_{3}=x_{4}=0\}$ & {} \\ \hline
%
{} & $ e_{1}$, 
& $\{ x_{1}=0 \}\cup $ & {}\\
 IV    & 
  $e_{3}$, 
& $\{ x_{6}=0 \}\cup \{ x_2=x_3= $ & $ {\Bbb P}_{(1,1,1,n)} $              \\
   {}  & $(n+1)e_{1}+ne_{2}+e_{3} $ 
& $x_{4}=x_{5}=0\}$ & {}   \\ \hline
%
{}    & $ e_{3}$,
 & $\{ x_{1}=x_{5}=0 \}\cup $ & {}\\
V     &
 $(n+1)e_{1}+ne_{2}+e_{3}$,   
 & $\{ x_{6}=0 \}\cup $ & singular toric    \\
    {} & $(n+1)e_{1}+(n+1)e_{2}+e_{3}$ 
  & $\{x_{2}=x_{3}=x_{4}=0\}$ & {}     \\        \hline
%
{}    & $e_{2}$, 
& $\{ x_{2}=0 \}\cup $ & {}\\
 VI    & 
  $e_{3}$, 
& $\{ x_{6}=0 \}\cup \{ x_1=x_3=$ & $ {\Bbb P}_{(1,1,1,n+1)} $              \\
   {}  & $(n+1)e_{1}+(n+1)e_{2}+e_{3} $ 
  & $x_{4}=x_{5}=0\}$ & {}   \\ \hline
\end{tabular}
\end{flushleft}
\end{table} \\
%
The contents of the phase transitions are as follows.
\begin{itemize}
\item %
On the phase boundary between I  and II,
the divisor %
$\{x_{1}=0 \}$ \\ %
$\cong$ ${\Bbb F}_{n+1}(x_3,x_4;x_6,x_2)$
of ${\Bbb H}_{n}$
contracts to the base ${\Bbb P}^{1}(x_3,x_4)$.
\item %
On the phase boundary between I and III,
the divisor  $\{x_{2}=0 \}$ \\ %
$\cong $ ${\Bbb F}_{n}(x_3,x_4;x_5,x_1)$ %
of ${\Bbb H}_{n}$
contracts to the base ${\Bbb P}^{1}(x_3,x_4)$.
\item %
On the phase boundary between  I and V,
the divisor $\{x_6=0\}$ \\ %
$\cong$ ${\Bbb P}^{2}(x_1,x_3,x_4)$ %
of ${\Bbb H}_{n}$
contracts to a point.
\item %
On the phase boundary between II and IV,
the divisor %
$\{ x_6=0 \}$ \\ %
$\cong$ ${\Bbb P}^{2}(x_2,x_3,x_4)$ 
of ${\Bbb G}_{n}$
contracts to a point.
\item %
On the phase boundary between III and VI,
the divisor %
$\{ x_6=0 \}$ \\ %
$\cong$  ${\Bbb P}^{2}(x_1,x_3,x_4)$ 
of ${\Bbb G}_{n+1}$
contracts to a point.
\end{itemize}
Recall that the gauge symmetry of
${\Bbb G}_{n}$ model is localized at the world volume of the
coincident seven-branes wrapped around the  divisor $\{ y_{2}=0 \}$.
The tension of the resulting three-brane
is proportional to the volume of the divisor \cite{MV2}.
Thus at  the phase boundary between II and IV (also between III and VI),
where the volume of the divisor becomes zero, 
the bare gauge  coupling diverses as described  in
the context of the strong coupling dual of the heterotic string realized 
by M theory on ${\Bbb S}^{1}/{\Bbb Z}_{2}$ $\times$ $Y_{3}$
\cite{Wi6,BD}. \\
%
Let us here describe the dual heterotic coupling 
and the gauge coupling 
in terms of the K\"ahler parameters of ${\Bbb G}_{n}$.\\
Define the divisors of ${\Bbb G}_{n}$ $d_{1}$, $d_{2}$ by
%
$
\{ x_{i}=0 \}=  d_{1},\ %
\{ y_{2}=0 \}=  d_{2}.  \\
$
Then $\{ y_{1}=0 \}$ is identified with $nd_{1}+d_{2}.$ \\
The intersection pairings of them \cite{MP} are  given by \\
$
d_{1}\cdot d_{1}\cdot d_{1} = 0, \ %
 d_{1}\cdot d_{1}\cdot d_{2}=1,  \ %
d_{1}\cdot d_{2}\cdot d_{2} =-n,\ %
d_{2}\cdot d_{2}\cdot d_{2}=n^2.\\
$
The Poincar\'e dual of the K\"ahler form is expressed as
$
\omega = (l_{1}+nl_{2}) d_{1}+l_{2}d_{2},
$
where $l_{2}$ is the radius of fiber %
${\Bbb P}^{1}_{\mbox{\scriptsize fiber}}(y)$
and $l_{1}$ is the radius of the exceptional  cross section
$\{ y_{2}=0 \}$ $\cong$ ${\Bbb P}^{2}(x)$.         \\
Then  volumes of toric cycles are evaluated as follows.
\begin{eqnarray}
\mbox{Vol}\left( {\Bbb P}^{1}_{\mbox{\scriptsize fiber}}\right) &=&
\omega \cdot d_{1} \cdot d_{1}=l_{2} \nonumber \\
\mbox{Vol}\left(\{y_{2}=0\} \right)&=& 
\frac{1}{2!}\ \omega \cdot \omega \cdot d_{2}
=\frac{1}{2} \ l_{1}^2      \\
\mbox{Vol}({\Bbb G}_{n})&=&
\frac{1}{3!} \  \omega \cdot \omega \cdot\omega
=\frac{1}{2}\left(  l_{1}^{2}+nl_{1}l_{2}
+\frac{1}{3}(nl_{2})^{2} \right)l_{2}.
\nonumber
\end{eqnarray}
We see that the heterotic coupling is given by
\begin{equation}
\exp(-2\phi)=\frac{1}{2}\left(  l_{1}^{2}+nl_{1}l_{2}
+\frac{1}{3}(nl_{2})^{2} \right).
\end{equation}
%
The phase boundary between II and IV is at $l_{1}=0$,
where the gauge coupling $g$ $\propto $ $1/l_{1}$
diverges \cite{Wi6,BD,MV2}.
The phase I could be regarded as a non-perturbative 
heterotic vacuum \cite{DMW,SW}
where a source term of   the Bianchi identity 
for 3-form field strength 
$H_{(3)}$ comes from a five-brane wrapped around 
a Riemann surface of $Y_3$ as pointed out in \cite{Wi6}. \\
Indeed we have two ${\Bbb P}^{1}$ fibration structures  in ${\Bbb H}_{n}$
over ${\Bbb P}^{2}_{\mbox{\scriptsize base}}$,
$\pi_{i}:{\Bbb H}_{n}\rightarrow {\Bbb P}^{2}_{\mbox{\scriptsize base}}$, 
$i=1,2$,  inherited from those of ${\Bbb G}_{n+1}$ and  
${\Bbb G}_{n}$.\\
The first ${\Bbb P}^{1}$ fibration is defined by
\begin{equation}
\pi_{1}(x_{1},...,x_{6})=(x_{1},x_{3},x_{4}).
\end{equation}
The generic fiber of $\pi_{1}$
is ${\Bbb P}^{1}$, while the one over the point on the divisor 
$\{ x_{1}=0 \} \subset {\Bbb P}^{2}_{\mbox{\scriptsize base}}$ 
is ${\Bbb P}^{1}\cup {\Bbb P}^{1}$. \\
The second ${\Bbb P}^{1}$ fibration is defined by
\begin{equation}
\pi_{2}(x_{1},...,x_{6})=(x_{2},x_{3},x_{4}).
\end{equation}
The fiber of $\pi_{2}$ over the divisor $\{ x_{2}=0 \}
\subset {\Bbb P}^{2}_{\mbox{\scriptsize base}}$ 
is ${\Bbb P}^{1}\cup {\Bbb P}^{1}$.\\
Thus we could say that  
IIB compactification on a  a ${\Bbb P}^{1}$ fiber bundle is dual to 
a perturbative heterotic string, while
one  on a ${\Bbb P}^{1}$ fibration which is not a fiber bundle 
to a non-perturbative heterotic string with five-branes.\\
In our example, we can see
the correspondence between F theory model and 
the dual heterotic string model more concretely.
The  second Chern class of the elliptic Calab-Yau threefold $Y_{3}$
is expressed as 
\begin{equation}
c_{2}(T_{Y_{3}})=36[\Sigma_{1}]+102[\Sigma_{2}], \label{secondChern}
\end{equation}
where $[\Sigma_{1}]$ represents the class of a rational curve
on the cross section of the elliptic fibration %
$\pi: Y_{3}\rightarrow {\Bbb P}^{2}$,
while $[\Sigma_{2}]$ the class of  a fiber elliptic curve. \\
Although we don't know the precise way of
dividing the second Chern class (\ref{secondChern}) above into those  of two
$\mbox{E}_{8}$ gauge bundles
in the heterotic string dual to ${\Bbb G}_{n}$ model,
we may conjecture that the second Chern classes of
two $\mbox{E}_{8}$ gauge bundles of 
the corresponding heterotic string  have  $(18\pm n) [\Sigma_{1}]$ 
as their components
and that the heterotic five-brane associated with
${\Bbb H}_{n}$ model is partially wrapped around a Riemann surface
$\Sigma_{1}$ on the cross section.
It would be interesting to further investigate $(0,2)$ heterotic string 
compactifications on $Y_{3}$ and elucidate the
relation to F theory models given here.

\newpage
\section{Flop Transitions}
\subsection{Blowing-Ups of ${\Bbb G}_{n}$ Along a Point}
\begin{figure}[h]
\psfig{file=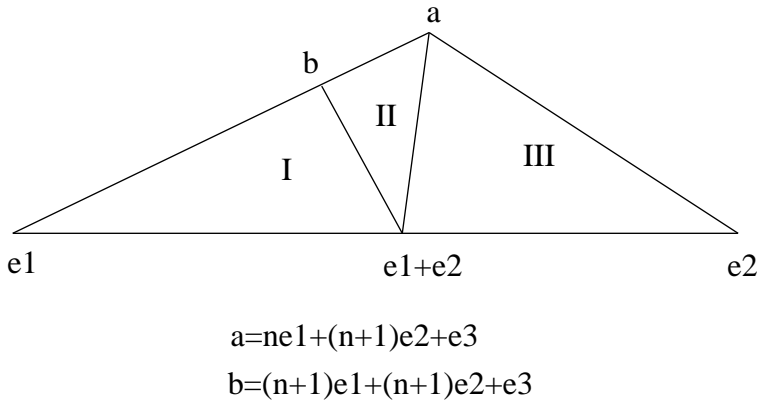,width=10cm}
\caption{ Phase Diagram of Exceptional Blowing-Up of ${\Bbb G}_{n}$}
\label{flop-phase}
\end{figure}

\begin{figure}[h]
\psfig{file=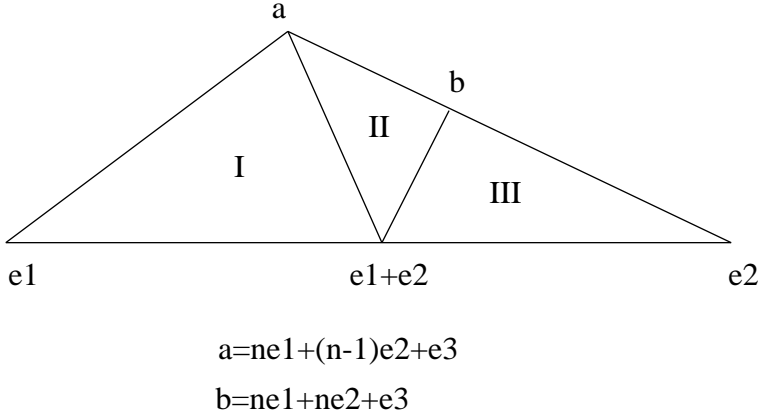,width=10cm}
\caption{ Phase Diagram of Normal Blowing-Up of ${\Bbb G}_{n}$}
\label{flopII-phase}
\end{figure}

In this section we present a few examples
of the model which admits the flop transition \cite{Wi4,AGM}
and joins a dual heterotic string vacuum
on  the  elliptic Calabi-Yau threefold
over ${\Bbb F}_{1}$ and that  over ${\Bbb P}^{2}$.\\
First let us take the toric threefolds defined by the data;
\begin{alignat}{6}
&x_{1}  & \quad %
&x_{2}  & \quad %
&x_{3}  & \quad %
&x_{4}  & \quad %
&x_{5}  & \quad %
&x_{6}       \nonumber \\ %
&1    & \quad %
&1    & \quad %
&1    & \quad %
&0    & \quad %
&n    & \quad %
&0         \nonumber \\  %
&0    & \quad %
&0    & \quad %
&1    & \quad %
&1    & \quad %
n&+1    & \quad %
&0         \nonumber \\ %
&0    & \quad %
&0    & \quad %
&0    & \quad %
&0    & \quad %
&1    & \quad %
&1,            %
\end{alignat}
We show only smooth phases in the phase diagram in %
Figure \ref{flop-phase}.
%
%
The phase I is the ${\Bbb P}^{1}(x_{5},x_{6})$ bundle over 
${\Bbb F}_{1}(x_{1},x_{2};x_{3},x_{4})$. \\
Acrossing the boundary wall between the phase I and II,
the 2-cycle \\    
${\Bbb P}^{1}(x_{1},x_{2})\cong \{ x_{4}=x_{5}=0\}$ \\ %
of the phase I is contracted to a point and replaced by the 2-cycle   \\%
${\Bbb P}^{1}(x_{4},x_{5})\cong \{ x_{1}=x_{2}=0\}$ \\ %
of the phase II.
Thus the phase II is a flop of the phase I. \\
As remarked in \cite{Wi1}, there appears the tensionless string carrying %
$N$ $=$ $2$ supersymmetry at the flop wall.
This string is obtained by wrapping the type IIB three-brane %
around the vanishing 2-cycle above.\\
On the boundary between the phase II and III,
the divisor \\ %
${\Bbb P}^{2}(x_{1},x_{2},x_{6})\cong \{ x_{4}=0 \}$ \\ %
of the phase II is blown down to a point.
The resulting threefold of the phase III
is precisely ${\Bbb G}_{n}$ treated in the previous section.
Thus the phase II is nothing but the 
blowing-up of ${\Bbb G}_{n}$ along the point 
$\{$ %
$ x_{1}$ $=$ $x_{2}$ $=$ $x_{6}$ $=0$ %
$\}$, %
which is contained in the exceptional cross section.\\
%
In the same way,
the blowing-up of ${\Bbb G}_{n}$ along the point
$\{$ %
$ x_{1}$ $=$ $x_{2}$ $=$ $x_{5}$ $=0$ %
$\}$, which is not contained in the exceptional cross section,
leads to the model with
the following toric data.
\begin{alignat}{6}
&x_{1}  & \quad %
&x_{2}  & \quad %
&x_{3}  & \quad %
&x_{4}  & \quad %
&x_{5}  & \quad %
&x_{6}       \nonumber \\ %
&1    & \quad %
&1    & \quad %
&1    & \quad %
&0    & \quad %
&n    & \quad %
&0         \nonumber \\  %
&0    & \quad %
&0    & \quad %
&1    & \quad %
&1    & \quad %
n&-1    & \quad %
&0         \nonumber \\ %
&0    & \quad %
&0    & \quad %
&0    & \quad %
&0    & \quad %
&1    & \quad %
&1,            %
\end{alignat}
There are  three smooth phases as shown in Figure \ref{flopII-phase}.
%
%
The phase I is a ${\Bbb P}^{1}$ bundle over ${\Bbb F}_{1}$ and 
the phase III is ${\Bbb G}_{n}$.
The phase boundary between I and II is a flopping  wall.
hence we have a tensionless string again there.\\
We have seen in this section that a blowing-up of ${\Bbb G}_{n}$ 
along a point gives a model  which admits a flop transition.
Its topology   depends on whether the point is in the exceptional 
cross section or not.\\
Note that the blowing-up of ${\Bbb G}_{n}$ around
a point treated here can be 
 regarded as ${\Bbb P}^{1}$ fibrations over ${\Bbb P}^{2}$,
where the fiber over the one point is ${\Bbb P}^{1} \cup {\Bbb P}^{2}$,
thus the phase II is also dual to a heterotic string.

%
\newpage
\section{${\Bbb P}^{2}$ bundle over ${\Bbb P}^{1}$ models}
Here we consider the toric threefold defined by the toric data
shown below,
\begin{alignat}{6}
&{} & \quad %
&x_{1} & \quad %
&x_{2} & \quad %
&y_{1} & \quad %
&y_{2} & \quad %
&y_{3} \nonumber \\  
&\lambda_{1} & \quad %
&1  & \quad %
&1  & \quad %
&n  & \quad %
&n  & \quad %
&0    \nonumber \\
&\lambda_{2}  & \quad %
&0  & \quad %
&0  & \quad %
&1  & \quad %
&1  & \quad %
&1          %
\end{alignat}
which we call ${\Bbb O}_{n}$. It has the structure of the ${\Bbb P}^{2}$
fiber bundle over ${\Bbb P}^{1}$ as;
\begin{equation}
\pi: {\Bbb O}_{n}(x,y)\longrightarrow {\Bbb P}^{1}(x).
\end{equation}
For $ 0 \leq n \leq 2$,
${\Bbb O}_{n}$ model has generically completely broken
non-Abelian  gauge symmetry, %
while for $2 \leq n \leq 12$ the Newton polytope of the elliptic 
Calabi-Yau fourfold over ${\Bbb O}_{n}$ is reflexive and 
isomorphic to that of \\
${\Bbb P}_{(1,1,n,n,4(n+1),6(n+1))}[12(n+1)]$,
which admits a transverse hypersurface for $n=$ 2,3,4,6,11 and 12.\\
We give the quantum numbers of models in Table \ref{On-models}.
\begin{table}[h]
\begin{center}
\caption{Quantum Numbers of ${\Bbb O}_{n}$ Models}
\label{On-models}
\begin{tabular}{|l||l|l|l|l|l|c|}\hline
{}& $h^{1,1}$ &$h^{1,3}$ &$h^{1,2}$ &$h^{2,2}$ &
$\chi(X_{4})/24$ & singularity type \\  \hline
${\Bbb O_{0}}$ & 3 & 3277 & 0 & 13164 & 822 & -\\  \hline
${\Bbb O_{1}}$ & 3 & 3277 & 0 & 13164 & 822 & -\\  \hline
${\Bbb O_{2}}$ & 3 & 3277 & 0 & 13164 & 822 & -\\  \hline
${\Bbb O_{3}}$ & 4 & 3443 & 7 & 13818 & 862 &
$IV^{ns}$ \\  \hline  %
${\Bbb O_{4}}$ & 5 & 3784 & 13 & 15174 & 946 &
$I_{0}^{*ns}$ \\  \hline %
${\Bbb O_{5}}$ & 7 & 4185 & 0 & 16812 & 1050 &
${IV^{*ns}}$ \\  \hline %
${\Bbb O_{6}}$ & 7 & 4613 & 12 & 18500 & 1154 &
$IV^{*ns}$ \\  \hline %
${\Bbb O_{7}}$ & 10 & 5056 & 0 & 20308 & 1268+2/4 &
${III^{*}}$ \\  \hline %
${\Bbb O_{8}}$ & 10 & 5514 & 0 & 22140 & 1383 &
${III^{*}}$ \\  \hline %
${\Bbb O_{9}}$ & 12 & 5972 & 24 & 23932 & 1492 &
$II^{*}$ \\  \hline %
${\Bbb O_{10}}$ & 12 & 6440 & 12 & 25828 & 1612 &
$II^{*}$ \\  \hline %
${\Bbb O_{11}}$ & 12 & 6908 & 0 & 27724 & 1732 &
$II^{*}$ \\  \hline %
${\Bbb O_{12}}$ & 24 & 7376 & 0 & 29644 & 1852 &
$II^{*}$ \\  \hline
\end{tabular}
\end{center}
\end{table}
%

%
\newpage
There is an identification:
${\Bbb O}_{1}$ $\cong$ ${\frak F}_{5}$.\\
The singularity type there  
means the one  at the divisor $\{ y_{3}=0 \}$ $\cong $ 
${\Bbb P}^{1}$ $\times$ ${\Bbb P}^{1}$.\\
Note that the rank of the vector multiplets 
$(h^{1,1}-3)$ predicted in (\ref{vector})
doesn't contradict with
the following identification of the non-Abelian gauge symmetries
localized at $\{ y_{3}=0 \}$:
$\mbox{G}_{2} \ (n=4)$, 
$\mbox{F}_{4} \ (n=5,6)$,
$\mbox{E}_{7} \ (n=7,8)$,
 $\mbox{E}_{8} \ (n=9,10,11,12)$,
which are obtained  from the singularity of the Weierstrass model.  
There is a remarkable similarity
between  the singularity of ${\Bbb O}_{n}$ model
and that of D=6 ${\Bbb F}_{n}$ model analyzed in \cite{MV2}.
We also remark that ${\Bbb O}_{n}$ model can be treated as a fiber-wise 
compactification \cite{KS}
to ${\Bbb P}^{1}_{\mbox{\scriptsize base}}$
of D=6 ${\Bbb P}^{2}_{\mbox{\scriptsize fiber}}$ model which has
the prescribed singularity at $\{ y_{3} = 0 \};$
namely the parameters of the  D=6  fiber theory 
cannot be chosen arbitrary.

\newpage
\section{${\Bbb P}_{1}$ Bundle over ${\Bbb F}_{a}$ Models}
\subsection{Model and Landau-Ginzburg Condition}
Here we consider the threefold ${\frak B}_{a,b,c}$ defined by 
the following toric data,
where $ (a,b,c) $ are positive integers,
\begin{alignat}{7}
&{}    & \quad %
&x_{1} & \quad %
&x_{2} & \quad %
&y_{1} & \quad %
&y_{2} & \quad %
&w_{1} & \quad %
&w_{2}   \nonumber \\
&\lambda_{1}  & \quad %
&1  & \quad %
&1  & \quad %
&a  & \quad %
&0  & \quad %
&b  & \quad %
&0  \nonumber \\
&\lambda_{2} & \quad %
&0  & \quad %
&0  & \quad %
&1  & \quad %
&1  & \quad %
&c  & \quad %
&0    \nonumber \\
&\lambda_{3} & \quad %
&0  & \quad %
&0  & \quad %
&0  & \quad %
&0  & \quad %
&1  & \quad %
&1      %
\end{alignat}
%
with the excluded set %
$E=\{x_1=x_2=0\}\cup\{y_1=y_2=0\}\cup\{w_1=w_2=0\}$. \\
%
As ${\frak B}_{a,b,c}$  has a structure of a ${\Bbb P}^{1}(w)$ bundle over
${\Bbb F}_{a}(x,y)$,
this  model is dual to the heterotic string
compactified on the elliptic Calabi-Yau threefold over
${\Bbb F}_{a}$. \\
Alternatively if we take the radius of ${\Bbb P}^{1}(x)$
 sufficiently large and regard
${\frak B}_{a,b,c}$  as a ${\Bbb F}_{c}(y,w)$ bundle 
over ${\Bbb P}^{1}(x)$,
this model can be analyzed as a 
fiberwise compactification  of D=6 model on ${\Bbb F}_{c}$ \cite{MV1,MV2}
on ${\Bbb P}^{1}$ as proposed in \cite{KS}.\\
%
%
The toric data for the elliptic Calabi-Yau fourfold over
${\frak B}_{a,b,c}$ realized by the Weierstrass model %
is as follows,
%
%
\begin{alignat}{9}
&x_{1} & \quad %
&x_{2} & \quad %
&y_{1} & \quad %
&y_{2} & \quad %
&w_{1} & \quad %
&w_{2} & \quad %
&z_{1} & \quad %
&z_{2} & \quad %
&z_{3}     \nonumber \\ %
%
&1   & \quad %
&1  & \quad %
&a  & \quad %
&0  & \quad %
&b  & \quad %
&0  & \quad %
&0  & \quad %
2(a+& b+2)    & \quad %
3(a+& b+2)    \nonumber \\
%
&0           & \quad %
&0           & \quad %
&1           & \quad %
&1           & \quad %
&c           & \quad %
&0           & \quad %
&0           & \quad %
2(c &+2)      & \quad %
3(c &+2)      \nonumber \\
%
&0           & \quad %
&0           & \quad %
&0           & \quad %
&0           & \quad %
&1           & \quad %
&1           & \quad %
&0           & \quad %
&4           & \quad %
&6           \nonumber \\
%
&0           & \quad %
&0           & \quad %
&0           & \quad %
&0           & \quad %
&0           & \quad %
&0           & \quad %
&1           & \quad %
&2           & \quad %
&3.                   %
\end{alignat}
%
We have observed  that the Newton polytope of the ${\frak B}_{a,b,c}$ model
is isomorphic to that of %
${\cal W}_{a,b}:={\Bbb P}_{(1,1,a,b,2(a+b+2),3(a+b+2))}[6(a+b+2)]$ if
\begin{eqnarray}
ac-b  \geq &0&      \nonumber  \\  
b-(a+2)  \geq &0&  \label{embedding}   \\
2b-c(a+2) \geq &0.& \nonumber
\end{eqnarray}
Thus the spectrum of the model satisfying (\ref{embedding}) can be computed 
by the Vafa formula of Landau-Ginzburg orbifolds \cite{Vacua}
if the corresponding ${\cal W}_{a,b}$
admits a transverse hypersurface.\\
Note that consequently the Newton polytopes %
of ${\frak B}_{a,b,c}$ model and  ${\frak B}_{a,b,c'}$ model
are isomorphic %
if both $(a,b,c)$ and $(a,b,c')$ satisfy the condition (\ref{embedding}).\\
For example let us take the case in which  
${\cal W}_{a,b}$ admits 
a Fermat type hypersurface.
In this case the facets of the Newton polytope 
of each  ${\frak B}_{a,b,c}$ model with $c$ 
satisfying (\ref{embedding}) is simplicial and  given by
\begin{eqnarray}
-m_{1} &\leq &  1, \nonumber \\ %
-m_{4} &\leq &  1, \nonumber \\ %
-m_{5} &\leq &  1, \nonumber \\
m_{3}+2(2m_{4}+3m_{5})& \leq & 1,  \nonumber \\
m_{2}-cm_{3}-(c-2)(2m_{4}+3m_{5})     & \leq & 1,  \nonumber \\
m_{1}-am_{2}+(ac-b)m_{3}+(ac-a-b+2)(2m_{4}+3m_{5})& \leq & 1. 
\end{eqnarray}
Upon the change of the variables over ${\Bbb Z}$:
\begin{eqnarray}
{m'}_{i} &=&  m_{i},\ \   i=1,4,5  \nonumber \\
-{m'}_{2} &=& m_{2}-cm_{3}-2(c-2)m_{4}-3(c-2)m_{5} \label{overZ}  \\
-{m'}_{3} &=& m_{3}+4m_{4}+6m_{5}, \nonumber
\end{eqnarray}
we obtain the well-known Newton polytope of ${\cal W}_{a,b}$:
\begin{eqnarray}
-{m'}_{1} &\leq &  1, \nonumber \\ %
-{m'}_{2} &\leq &  1, \nonumber \\ %
-{m'}_{3} &\leq &  1, \nonumber \\ %
-{m'}_{4} &\leq &  1, \nonumber \\ %
-{m'}_{5} &\leq &  1, \nonumber \\ %
{m'}_{1}+a{m'}_{2}+b{m'}_{3}+(a+b+2)(2{m'}_{4}+3{m'}_{5})& \leq & 1. 
\end{eqnarray}
%
%
%
For another example in favour of our conjecture,
take ${\frak B}_{10,48,c}$ models.
We have checked that each of ${\frak B}_{10,48,5}$, 
${\frak B}_{10,48,6}$, %
${\frak B}_{10,48,7}$, and ${\frak B}_{10,48,8}$
has a Newton polytope isomorphic to that of %
%
${\cal W}_{10,48}$:
the isomorphism between  the Newton polytope of ${\frak B}_{10,48,c}$
and that of ${\cal W}_{10,48}$
are given by the same form as (\ref{overZ}). \\ 
%
%
%
%
In passing we note that some of ${\frak B}_{a,b,c}$ models (e.g. 
${\frak B}_{4,7,2}$)  have non-reflexive Newton polytopes.
It would be interesting if we can classify all the ${\frak B}_{a,b,c}$
models that  have reflexive Newton polytopes.

\newpage
\subsection{Quantum Numbers}
In this subsection we present the physical Hodge numbers
for some models.\\
First we treat  ${\frak B}_{3,11,4}$ model.
This model has been analyzed as a model of
$\mbox{Sp}_{1}$ gluino condensation using
fiberwise compactification technique of  dual 
D=6 heterotic string which has generically 
an $\mbox{SO}_{32}$ zero-size instanton \cite{KS}.\\
The Newton polytope of the ${\frak B}_{3,11,4}$ 
model, which is  %
isomorphic to that of ${\cal W}_{3,11}$ model, is reflexive  
with nine facets and eleven vertexes  
and thus  the physical Hodge numbers can be computed 
by the Batyrev formula (\ref{batyrev}),
\begin{equation}
(h^{1,1},h^{3,1},h^{2,1},h^{2,2})
=(10,7497,0,30072).
\end{equation}
Second we list a few examples of the spectra in Table \ref{Babc-models}
which can be computed by the Vafa formula
using the above-mentioned isomorphism of the Newton polytopes.
\begin{table}[h]
\begin{center}
\caption{Quantum Numbers of ${\frak B}_{a,b,c}$ Models}
\label{Babc-models}
\begin{tabular}{|l||l|l|l|l|l|}\hline
$(a,b,c)$ & %
$h^{1,1}$ &  %
$h^{1,3}$ &  %
$h^{1,2}$ &  %
$h^{2,2}$ &  %
$\chi(X_{4})/24$ \\ %
 \hline              %
$(2,8,2)$ &    8 & 6528 & 0 & 26188 & 822\\  \hline %
$(3,10,4)$  & 9  & 6796   & 1 & 27262 & 1703 \\  \hline %
$(3,15,5)$ & 12 & 10727 & 3 & 42994 & 2686 \\  \hline   %
$(3,27,10)$ & 16 & 24371 & 3 & 97586 & 6098 \\  \hline %
$(3,30,10)$ & 15 & 28692 & 3 & 114866 & 7178 \\  \hline %
$(4,36,9)$  & 18  & 30988 & 6 & 124056 & 7752 \\ \hline %
$(5,14,3)$ & 13 & 7991 & 0 & 32060 & 2003 \\ \hline %
$(6,16,4)$  & 16 & 8698   & 2 & 34896 & 2180 \\  \hline %
$(7,11,2)$ & 11 & 6665 & 3 & 26742 & 1670+1/4   \\ \hline %
$(8,60,8)$ & 30 & 43037 & 3 & 172306 & 10768 \\ \hline %
$(9,13,2)$ & 12 & 7144  & 0 & 28668 & 1791  \\  \hline  %
$(10,18,3)$& 18 & 9066  & 4 & 36372 & 2272 \\   \hline  %
$(11,77,7)$ & 43 & 51837 & 0 & 207564 & 12972 \\ \hline %
$(12,36,3)$& 32 & 17464 & 0 & 70028 & 4376 \\ \hline    %
\end{tabular}
\end{center}
\end{table}\\
We remark that ${\frak B}_{6,16,4}$ model has been treated in \cite{KS}
as a model of  world sheet instanton destabilization,
and the resolution of ${\cal W}_{2,8}$ model has been given  in \cite{BS}.
%
\newpage
\section{Superpotentials}
\subsection{Divisor and its Normal Bundle}
In this section we investigate in some detail 
the generation of superpotential
due to type IIB three-brane wrapped around the divisor
$W_{2}:=\pi(D_{3})$ of $B_{3}$.
We shall confine our attention  to such  divisors  that
\begin{equation}
\pi: D_{3} \longrightarrow W_{2}=\pi(D_{3}) \label{elliptic-divisor},
\end{equation}
which is a divisor of $X_{4}$,
is a smooth elliptic  threefold.\\
As the superpotential is generated by wrapping a three-brane
 around $W_{2}$,
we expect  that its  effect is  depend only on 
the neighbourhood of $W_{2}$ in $B_{3}$  i.e., 
the normal bundle ${\cal N}_{W_{2}}$ of $W_{2}$ in $B_{3}$.\\
In fact the elliptic threefold
$D$ in (\ref{elliptic-divisor})
is determined solely by the normal bundle ${\cal N}_{W_{2}}$.\\
Here let us take ${\Bbb G}_{n}$ model for example.
It has a distinguished divisor;
\begin{equation}
W_{2}:=\{ y_{2}=0 \}\cong {\Bbb P}^{2}(x),
\end{equation} 
which is the exceptional cross section.  \\
The normal bundle of this divisor is
\begin{equation}
{\cal N}_{W_{2}}\cong {\cal O}_{{\Bbb P}^{2}}(-n),
\end{equation}
and it is described as a non-compact toric threefold
with the following data:
\begin{alignat}{4}
&x_{1}  & \quad   %
&x_{2}  & \quad   %
&x_{3}  & \quad   %
&f       \nonumber \\ %
&1   & \quad %
&1   & \quad %
&1   & \quad %
-&n           %
\end{alignat}
where $f$ is the fiber coordinate and 
the excluded set is $\{ x_{1}=x_{2}=x_{3}=0 \}$.\\
Restricting  the Weierstrass model over the non-compact threefold %
${\cal O}_{W_{2}}(-n)$ to the zero section $\{ f=0 \}$,
 we arrive at the following toric data for 
 the elliptic threefold $D_{3}$ over $W_{2}$,             %
\begin{alignat}{6}
&x_{1}  & \quad  %
&x_{2}  & \quad  %
&x_{3}  & \quad  %
&z_{1}  & \quad  %
&z_{2}  & \quad  %
&z_{3}     \nonumber \\ %
&1   & \quad  %
&1   & \quad  %
&1   & \quad  %
&0   & \quad  %
2(3&-n)   & \quad  %
3(3&-n)     \nonumber \\ %
&0   & \quad  %
&0   & \quad  %
&0   & \quad  %
&1   & \quad  %
&2   & \quad  %
&3            %
\end{alignat}  
This threefold is smooth only for $n\leq 3,$
and we  deal with only such  cases.\\
The generalization of the above argument to any pair 
$(W_{2},{\cal N}_{W_{2}})$
of the surface and its normal bundle is straightforward.\\
The elliptic threefold %
$\pi: D_{3} \rightarrow W_{2}$ 
is described by 
the following Weierstrass model over $W_{2}$,
\begin{equation}
(z_{3})^{2}= (z_{2})^{3}+ z_{2}(z_{1})^{4}F(w) + (z_{1})^{6}G(w),%
\label{Weierstrass/W}  
\end{equation}
where
%
$F \in \Gamma(-4K_{W_{2}}+4[{\cal N_{W_{2}}}])$, \quad %
$G \in \Gamma(-6K_{W_{2}}+6[{\cal N_{W_{2}}}])$.\\
%
%
Instead of discussing  the general theory, %
it is appropriate  here to treat the surface   $ W_{2}={\Bbb F}_{n}$,
which often appears as a toric divisor.\\
Let us define the line bundle ${\cal O}_{{\Bbb F}_{n}}(l[F]+m[E])$
on ${\Bbb F}_{n}$ 
 by the following toric data,
%
\begin{alignat}{5}
&x_{1} & \quad  %
&x_{2} & \quad  %
&y_{1} & \quad  %
&y_{2} & \quad  %
&f      \nonumber \\  %
&1  & \quad   %
&1  & \quad   %
&n  & \quad   %
&0  & \quad   %
&l  \nonumber \\ %
&0  & \quad   %
&0  & \quad   %
&1  & \quad   %
&1  & \quad   %
&m            \label{Fn-Normal}%
\end{alignat}
where $[F]$ represents the fiber divisor $\{  x_{i}=0 \},$
and $[E]$ represents the exceptional 
cross section divisor $\{  y_{2}=0 \}$
of ${\Bbb F}_{n}$ \cite{Oda}.\\
If the normal bundle of $W$ $=$ ${\Bbb F}_{n}$ is isomorphic to
${\cal O}_{{\Bbb F}_{n}}(l[F]+m[E])$, %
then the elliptic threefold %
$\pi: D_{3}\rightarrow W_{2}$ %
is given by the Weierstrass model over 
${\Bbb F}_{n}$ where the charges  of  $F$ and $G$
with respect to (\ref{Fn-Normal})   are  $(4(l+n+2),4(m+2))$ and  
$(6(l+n+2),6(m+2))$ respectively.\\ %
We present the toric data of  the Weierstrass model 
in this case for convenience;
\begin{alignat}{7}
&x_{1}  & \quad %
&x_{2}  & \quad %
&y_{1}  & \quad %
&y_{2}  & \quad %
&z_{1}  & \quad %
&z_{2}  & \quad %
&z_{3}  \nonumber  \\ %
&1  & \quad %
&1  & \quad %
&n  & \quad %
&0  & \quad %
&0  & \quad %
2(n+&l+2)  & \quad %
3(n+&l+2)  \nonumber \\ %
&0  & \quad %
&0  & \quad %
&1  & \quad %
&1  & \quad %
&0  & \quad %
2(m&+2)  & \quad %
3(m&+2)  \nonumber \\ %
&0  & \quad %
&0  & \quad %
&0  & \quad %
&0  & \quad %
&1  & \quad %
&2  & \quad %
&3          %
\end{alignat}
By the investigation of the equation (\ref{Weierstrass/W}),
we see that the elliptic fibration \\ %
$\pi:$ $ D_{3}$ $\rightarrow$ $W_{2}$ 
is generically  smooth 
 if
\begin{equation}
(l+n+2)-n(m+2) \geq  0, \ \mbox{and}\quad   m+2 \geq  0.
\label{smoothD} 
\end{equation}
We deal with only such cases. For the case of singular divisors
see \cite{KV-BJPSV}.\\
A divisor is called {\it exceptional}
if it  can be contracted without leaving 
any singularities on $B_{3}$.
The normal bundle of an exceptional divisor ${\Bbb F}_{n}$ 
is of type $ m = -1 $.
In fact, under the blowing-down
the divisor contracts to ${\Bbb P}^{1}$ as follows,
\begin{equation}
({\Bbb F}_{n},{\cal O}_{{\Bbb F}_{n}}(l[F]-[E])) %
\longrightarrow %
({\Bbb P}^{1},{\cal O}_{{\Bbb P}^{1}}(l)\oplus %
{\cal O}_{{\Bbb P}^{1}}(n+l)). 
\end{equation}

\subsection{Detailed Analysis of Superpotential}
Recall that we have presented  the superpotential anomaly
as an intersection pairing on $B_{3}$ in (\ref{superpot3}).
However once we know the normal bundle of $W_{2}$ in $B_{3}$,
we can further reduce it to a intersection pairing 
on the surface  $W_{2}$ as follows,
\begin{equation}
\chi(D_{3},{\cal O}_{D_{3}})=-\frac{1}{2}[{\cal N}_{W_{2}}]\cdot %
(c_{1}(W_{2})+[{\cal N}_{W_{2}}]).\label{superpot2}
\end{equation}
For the pair $({\Bbb P}^{2}, {\cal O}_{{\Bbb P}^{2} }(-n))$,
 (\ref{superpot2}) gives
\begin{equation}
\chi(D_{3},{\cal O}_{D_{3}})=\frac{1}{2}n(3-n).
\end{equation}
There are the two cases $n=1$ and  $n=2$
that satisfy both the necessary condition \cite{Wi2} 
for the superpotential generation: 
$
\chi(D_{3},{\cal O}_{D_{3}})=1,
$
and the smoothness
condition: $n\leq 3$ of the previous subsection. \\
According to \cite{Grassi},
there are the  following cohomological relations 
for any smooth pair $(D_{3},W_{2})$ 
of an elliptic fibration (\ref{smoothD});
\begin{eqnarray}
H^{0}(D_{3},{\cal O}_{D_{3}})&\cong &  H^{0}(W_{2},{\cal O}_{W_{2}}), 
 \nonumber \\ %
0\rightarrow H^{1}(W_{2},{\cal O}_{W_{2}})
&\rightarrow & H^{1}(D_{3},{\cal O}_{D_{3}})
\rightarrow H^{2}(W_{2},{\cal N}_{W_{2}})^{*}\rightarrow 0, 
\ \ (\mbox{exact})  \nonumber \\ %
0\rightarrow H^{2}(W_{2},{\cal O}_{W_{2}})
&\rightarrow & H^{2}(D_{3},{\cal O}_{D_{3}})
\rightarrow H^{1}(W_{2},{\cal N}_{W_{2}})^{*}\rightarrow 0, 
\ \ (\mbox{exact})  \nonumber  \\ %
H^{3}(D_{3},{\cal O}_{D_{3}})&\cong &  H^{0}(W_{2},{\cal N}_{W_{2}})^{*} %
\label{D-cohomology} %
\end{eqnarray}
Using  these relations, we can see  that  both $n=1$ and $n=2$ cases
satisfy the sufficient condition for superpotential generation:
$
h^{i,0}(D_{3})=0, \quad i=1,2,3.
$
For example, both ${\Bbb G}_{1}$ model  and  ${\Bbb G}_{2}$ model
have  a  superpotential. \\
%
As for  the pair $({\Bbb F}_{n},{\cal O}(l[F]+m[E]))$, (\ref{superpot2}) yields
\begin{equation}
\chi(D_{3},{\cal O}_{D_{3}})=1-(1+m)(1+l)+\frac{1}{2}m(1+m)n.
\end{equation}
The pair $({\Bbb F}_{n},{\cal O}_{{\Bbb F}_{n}}(l[F]+m[E]))$
 satisfies both the smoothness condition (\ref{smoothD}) and
the necessary condition 
$ \chi(D_{3},{\cal O}_{D_{3}}) =1 $ if and only if  %
\begin{equation}
m = -1, \quad \mbox{and} \quad l+2 \geq 0. \label{aiu}
\end{equation}
We can show  using (\ref{D-cohomology}) that
any  divisor which satisfies the condition (\ref{aiu})
gives rise to the superpotential.
Thus  all the exceptional  divisors ${\Bbb F}_{n}$
appeared in the blowing-up/down transitions
of the Fano threefolds in Figure \ref{18Fanos}
contribute to the superpotential.
\subsection{Change in Quantum Numbers by Blowing-Down}
%
By computing the change in intersection pairing  $c_{1}(B)^{3}$, 
we obtain the following formula for the increase of the Euler number
under the blowing-down of ${\Bbb F}_{n}$ with 
the  normal bundle of type $(-1,l)$ with $l\geq -2$:
\begin{equation}
\frac{1}{24}\varDelta \chi=30(3+n+2l). \label{number3-brane}
\end{equation}
At least for smooth models this formula should be valid.
We may also conjecture that
the validity of (\ref{number3-brane}) in general as blowing-downs are local
operations on threefolds.
Then,  under the assumption: $ \varDelta h^{1,1}=-1$,
the Hodge numbers of the Calabi-Yau fourfold change as
\begin{eqnarray}
\varDelta h^{1,3} &=& 1+120(3+n+2l)+\varDelta h^{1,2}, \nonumber \\
\varDelta h^{2,2} &=& 480(3+n+2l)
+2\varDelta h^{1,2}. \label{changeinHodgenumbers}    
%
%
%
\end{eqnarray}
Unfortunately $\varDelta h^{1,2}$ above is not always zero.\\
A  counter-example is the following:
${\Bbb H}_{2}\rightarrow {\Bbb G}_{2}$,
which is associated with the exceptional divisor 
${\Bbb F}_{3}$ of type $(-1,-2)$ and
has $\varDelta h^{1,2}=-1$.   \\
We present for completeness the analogous formula for 
a blowing-down of exceptional ${\Bbb P}^{2}$ to a point:
\begin{eqnarray}
\frac{1}{24}\varDelta\chi &=& 30\cdot 4 \nonumber \\
\varDelta h^{1,3} &=& 1+120\cdot 4+\varDelta h^{1,2}, \nonumber \\
\varDelta h^{2,2} &=& 480 \cdot 4+2\varDelta h^{1,2}.  \label{P2}
%
%
%
\end{eqnarray}
Indeed the quantum numbers of the toric Fano models ${\frak F}_{n}$
shown in Table 2 are in accord with the formulas (\ref{number3-brane}), 
(\ref{changeinHodgenumbers}) and (\ref{P2})
with $\varDelta h^{1,2}=0$. \\
We  have also checked that  the formula (\ref{changeinHodgenumbers})
with
$\Delta h^{1,2}=0$  can be used to the blowing-down: 
$
{\Bbb H}_{n}\rightarrow {\Bbb G}_{n+1},
$
where the exceptional divisor is ${\Bbb F}_{n}$ of type $(-1,1)$,
which satisfies (\ref{aiu}).
Namely  the  physical Hodge numbers of the
${\Bbb H}_{n}$ model is known  from those of     
 ${\Bbb G}_{n+1}$ model.

\vspace{2cm}
{\it Acknowledgement}\\
The author would like to thank the members of KEK
for  discussions.

\newpage
\appendix
\section{Toric data for Fano Threefolds}
We give the toric data of the 
Fano threefolds.
The toric data of ${\frak F}_{14}$ is as follows.
\begin{alignat}{7}
&x_{7}  & \quad %
&x_{8}  & \quad %
&x_{2}  & \quad %
&x_{4}  & \quad %
&x_{6}  & \quad %
&x_{5}  & \quad %
&x_{1}  \nonumber \\
&1  & \quad %
&1  & \quad %
&0  & \quad %
&0  & \quad %
&0  & \quad %
&0  & \quad %
-&1     \nonumber \\ %
&0  & \quad %
&0  & \quad %
&1  & \quad %
&0  & \quad %
&1  & \quad %
&0  & \quad %
-&1  \nonumber \\ %
&0  & \quad %
&0  & \quad %
&0  & \quad %
&1  & \quad %
&1  & \quad %
-&1  & \quad %
&0 \nonumber \\ %
&0  & \quad %
&0  & \quad %
&0  & \quad %
&0  & \quad %
-&1  & \quad %
&1  & \quad %
&1      %
\end{alignat}
The toric data of ${\frak F}_{15}$ is as follows.
\begin{alignat}{7}
&x_{7}  & \quad %
&x_{8}  & \quad %
&x_{1}  & \quad %
&x_{3}  & \quad %
&x_{5}  & \quad %
&x_{4}  & \quad %
&x_{6}  \nonumber \\
&1  & \quad %
&1  & \quad %
-&1  & \quad %
&0  & \quad %
&0  & \quad %
&0  & \quad %
&0  \nonumber \\
&0  & \quad %
&0  & \quad %
&1  & \quad %
&0  & \quad %
&1  & \quad %
&0  & \quad %
-&1 \nonumber \\
&0  & \quad %
&0  & \quad %
&0  & \quad %
&1  & \quad %
&1  & \quad %
-&1  & \quad %
&0  \nonumber \\
&0  & \quad %
&0  & \quad %
&0  & \quad %
&0  & \quad %
-&1  & \quad %
&1  & \quad %
&1         %
\end{alignat}
The toric data of ${\frak F}_{16}$ is as follows.
\begin{alignat}{7}
&x_{7}  & \quad %
&x_{8}  & \quad %
&x_{3}  & \quad %
&x_{5}  & \quad %
&x_{1}  & \quad %
&x_{6}  & \quad %
&x_{2}  \nonumber \\
&1  & \quad %
&1  & \quad %
&0  & \quad %
&0  & \quad %
-&1  & \quad %
&0  & \quad %
&0  \nonumber \\
&0  & \quad %
&0  & \quad %
&1  & \quad %
&0  & \quad %
&1  & \quad %
&0  & \quad %
-&1 \nonumber \\
&0  & \quad %
&0  & \quad %
&0  & \quad %
&1  & \quad %
&1  & \quad %
-&1  & \quad %
&0  \nonumber \\
&0  & \quad %
&0  & \quad %
&0  & \quad %
&0  & \quad %
-&1  & \quad %
&1  & \quad %
&1         %
\end{alignat}
Each of above three models has a structure of a 
non-trivial ${\Bbb S}_{2}$ bundle over ${\Bbb P}^{1}$.\\
The toric data of ${\frak F}_{13}\cong
{\Bbb P}^{1} \times {\Bbb S}_{2}$ is given by
\begin{alignat}{7}
&x_{1}  & \quad %
&x_{2}  & \quad %
&x_{3}  & \quad %
&x_{4}  & \quad %
&x_{5}  & \quad %
&x_{6}  & \quad %
&x_{7}      \nonumber \\ %
&1      & \quad %
&1      & \quad %
&0      & \quad %
&0      & \quad %
&0      & \quad %
&0      & \quad %
&0      \nonumber \\ %
&0      & \quad %
&0      & \quad %
&0      & \quad %
&1      & \quad %
&1      & \quad %
&0      & \quad %
-&1      \nonumber \\ %
&0      & \quad %
&0      & \quad %
&1      & \quad %
&0      & \quad %
&1      & \quad %
-&1      & \quad %
&0        \nonumber \\ %
&0      & \quad %
&0      & \quad %
&0      & \quad %
&0      & \quad %
-&1      & \quad %
&1      & \quad %
&1         %
\end{alignat}
The toric data of ${\frak F}_{12}$ model is
\begin{alignat}{7}
 &{}         &\quad %
   &x_{1}   & \quad %
    &x_{2}  & \quad %
&x_{3} & \quad %
&x_{4} & \quad %
 &x_{5} & \quad %
 &x_{6}    \nonumber \\ %
&{} & \quad %
&1     & \quad %
&1     & \quad %
&1     & \quad %
&0     &\quad  %
  -&1   & \quad %
&0     \nonumber \\ %
&{} & \quad %
 -&1     & \quad %
 &0     & \quad %
&0     & \quad  %
&0     &   \quad %
 &1   & \quad %
 &1      \nonumber \\     %
&{} &\quad %
  &1     & \quad %
&0     & \quad %
&0     & \quad  %
&1     & \quad %
   &0   & \quad %
 -&1.         %
\end{alignat}
%
The toric data of ${\frak F}_{8}$ model is
\begin{alignat}{6}
&x_{1 }  & \quad %
&x_{2 }  & \quad %
&x_{3 }  & \quad %
&x_{4 }  & \quad %
&x_{5 }  & \quad %
&x_{6 }  \nonumber \\
&1  & \quad %
&1  & \quad %
&0  & \quad %
&0  & \quad %
&0  & \quad %
-&1    \nonumber \\ %
&0  & \quad %
&0  & \quad %
&1  & \quad %
&1  & \quad %
-&1  & \quad %
&0     \nonumber \\ %
&0  & \quad %
&0  & \quad %
&0  & \quad %
&0  & \quad %
&1  & \quad %
&1   %
\end{alignat}
The toric data of ${\frak F}_{7}$ is given by
\begin{alignat}{6}
&x_{1 }  & \quad %
&x_{2 }  & \quad %
&x_{3 }  & \quad %
&x_{4 }  & \quad %
&x_{5 }  & \quad %
&x_{6 }  \nonumber \\
&1  & \quad %
&1  & \quad %
&0  & \quad %
&0  & \quad %
&0  & \quad %
-&1    \nonumber \\ %
&0  & \quad %
&0  & \quad %
&1  & \quad %
&1  & \quad %
&0  & \quad %
-&1     \nonumber \\ %
&0  & \quad %
&0  & \quad %
&0  & \quad %
&0  & \quad %
&1  & \quad %
&1   %
\end{alignat}
%
Both ${\frak F}_{7}$ and ${\frak F}_{8}$ are 
non-trivial ${\Bbb P}^{1}$
bundles over ${\Bbb P}^{1}\times {\Bbb P}^{1}$.\\
The toric data of ${\frak F}_{9}$  is given by
\begin{alignat}{6}
&x_{1 }  & \quad %
&x_{2 }  & \quad %
&x_{3 }  & \quad %
&x_{4 }  & \quad %
&x_{5 }  & \quad %
&x_{6 }  \nonumber \\
&1  & \quad %
&1  & \quad %
&0  & \quad %
-&1  & \quad %
&0  & \quad %
&0    \nonumber \\ %
&0  & \quad %
&0  & \quad %
&1  & \quad %
&1  & \quad %
&0  & \quad %
&0     \nonumber \\ %
&0  & \quad %
&0  & \quad %
&0  & \quad %
&0  & \quad %
&1  & \quad %
&1   %
\end{alignat}
so that ${\frak F}_{9}$ is isomorphic to
${\Bbb P}^{1}\times {\Bbb F}_{1}$.\\
The toric data of ${\frak F}_{10}$ which is a 
${\Bbb P}^{1}$ bundle over ${\Bbb F}_{1}$ is given by
\begin{alignat}{6}
&x_{1}  & \quad %
&x_{2}  & \quad %
&x_{3}  & \quad %
&x_{4}  & \quad %
&x_{5}  & \quad %
&x_{6}  \nonumber \\ %
&1  & \quad %
&1  & \quad %
&0  & \quad %
-&1  & \quad %
&0  & \quad %
&0    \nonumber \\ %
&0  & \quad %
&0  & \quad %
&1  & \quad %
&1  & \quad %
&0  & \quad %
-&1   \nonumber \\ %
&0  & \quad %
&0  & \quad %
&0  & \quad %
&0  & \quad %
&1  & \quad %
&1         %
\end{alignat}


\end{document}